\documentclass[iop,revtex4]{emulateapj}
\pdfoutput=1

\usepackage{graphicx}
\usepackage{natbib}
\usepackage{relsize}
\usepackage{textcomp}
\usepackage{upgreek}
\usepackage{amsmath}
\usepackage{amssymb}
\usepackage{calrsfs}

\setcitestyle{notesep={}}
\newcommand\nar{\ref@jnl{New A Rev.}}

\newcommand{\cii}{C\ensuremath{\,\textsc{ii}}}
\newcommand{\ciii}{C\ensuremath{\,\textsc{iii}}}
\newcommand{\ha}{H\ensuremath{\alpha}}
\newcommand{\lya}{Ly\ensuremath{\alpha}}

\newcommand{\heii}{He\ensuremath{\,\textsc{ii}}}
\newcommand{\civ}{C\ensuremath{\,\textsc{iv}}}

\def\lsim{\mathrel{\rlap{\lower 3pt \hbox{$\sim$}} \raise 2.0pt \hbox{$<$}}}
\def\gsim{\mathrel{\rlap{\lower 3pt \hbox{$\sim$}} \raise 2.0pt \hbox{$>$}}}

\def\lsun{{\rm\,L_\odot}}
\def\msun{{\rm M}_\odot}
\def\msunyr{{\rm M}_\odot\,{\rm yr}^{-1}}

\def\J0305{\mbox{J0305$-$3150}}

\def\flya{{\rm F}({\rm Ly}\alpha)}
\def\lfir{{\rm L}_{\rm FIR}}
\def\ltir{{\rm L}_{\rm TIR}}
\def\llya{{\rm L}({\rm Ly}\alpha)}
\def\lha{{\rm L}({\rm H}\alpha)}
\def\mbh{{\rm M}_{\rm BH}}
\def\M1450{{\rm M}_{\rm 1450}}

\shorttitle{Mapping the Lyman-Alpha Emission Around J0305$-$3150}

\shortauthors{Farina et al.}

\begin{document}

\title{Mapping the Lyman-Alpha Emission Around a $z$$\sim$6.6 QSO with MUSE:\\
Extended Emission and a Companion at Close Separation}

\author{
Emanuele~P.~Farina\altaffilmark{1},
Bram~P.~Venemans\altaffilmark{1},
Roberto~Decarli\altaffilmark{1},
Joseph~F.~Hennawi\altaffilmark{1},
Fabian~Walter\altaffilmark{1,2,3},
Eduardo~Ba{\~n}ados\altaffilmark{4,5},
Chiara~Mazzucchelli\altaffilmark{1},
Sebastiano~Cantalupo\altaffilmark{6},
Fabrizio~Arrigoni--Battaia\altaffilmark{7}, and
Ian~D.~McGreer\altaffilmark{8}\\
}
\altaffiltext{1}{Max Planck Institut f\"ur Astronomie, K\"onigstuhl~17, D-69117, Heidelberg, Germany}
\altaffiltext{2}{California Institute of Technology, Astronomy Department, MC105-24, Pasadena, CA-91101, United States of America} 
\altaffiltext{3}{National Radio Astronomy Observatory, Pete V. Domenici Array Science Center, P.O. Box O, Socorro, NM-87801, United States of America} 
\altaffiltext{4}{The Observatories of the Carnegie Institute of Washington, 813~Santa Barbara Street, Pasadena, CA-91101, United States of America} 
\altaffiltext{5}{Carnegie--Princeton Fellow}
\altaffiltext{6}{Institute for Astronomy, Department of Physics, ETH Z\"urich, Wolfgang--Pauli--Strasse~27, CH-8093, Z\"urich, Switzerland}
\altaffiltext{7}{European Southern Observatory, Karl--Schwarzschild--Str.~2, D-85748, Garching bei M\"unchen, Germany}
\altaffiltext{8}{Steward Observatory, 933~North Cherry Avenue, Tucson, AZ-85721, United States of America} 
\email{emanuele.paolo.farina@gmail.com}

\begin{abstract}
We utilize the Multi Unit Spectroscopic Explorer (MUSE) on the Very Large Telescope (VLT) to search for extended \lya\ emission around the $z$$\sim$6.6 QSO \J0305.
After carefully subtracting the point--spread--function, we reach a nominal 5--$\sigma$ surface brightness limit of SB$_{5\sigma}$=1.9$\times$10$^{-18}$\,erg\,s$^{-1}$\,cm$^{-2}$\,arcsec$^{-2}$ over a 1\,arcsec$^2$ aperture, collapsing 5~wavelength slices centered at the expected location of the redshifted \lya\ emission (i.e. at 9256\,\AA).
Current data suggest the presence (5--$\sigma$, accounting for systematics) of a \lya\ nebula that extends for 9\,kpc around the QSO.
This emission is displaced and redshifted by 155\,km\,s$^{-1}$ with respect to the location of the QSO host galaxy traced by the [\cii]\,158\,$\mu$m emission line.
The total luminosity is $\llya$=(3.0$\pm$0.4)$\times$10$^{42}$\,erg\,s$^{-1}$.
Our analysis suggests that this emission is unlikely to rise from optically thick clouds illuminated by the ionizing radiation of the QSO.
It is more plausible that the \lya\ emission is due to fluorescence of the highly ionized optically thin gas. 
This scenario implies a high hydrogen volume density of $n_{\rm H}$$\sim$6\,cm$^{-3}$.
In addition, we detect a \lya\ emitter (LAE) in the immediate vicinity of the QSO: i.e., with a projected separation of $\sim$12.5\,kpc and a line--of--sight velocity difference of 560\,km\,s$^{-1}$.
The luminosity of the LAE is $\llya$=(2.1$\pm$0.2)$\times$10$^{42}$\,erg\,s$^{-1}$ and its inferred star--formation--rate is SFR$\sim$1.3\,M$_\odot$\,yr$^{-1}$.
The probability of finding such a close LAE is one order of magnitude above the expectations based on the QSO--galaxy cross--correlation function.
This discovery is in agreement with a scenario where dissipative interactions favour the rapid build--up of super--massive black holes at early Cosmic times. 
\end{abstract}

\keywords{cosmology: observations -- galaxies: high-redshift -- quasars: general -- quasars: individual J0305$-$3150}

\vfil
\eject
\clearpage

\section{INTRODUCTION}\label{sec:introduction}

The study of QSOs at $z$$>$5.6 plays a central role in our understanding of how supermassive black holes (SMBH) and galaxies form in the early Universe.
Currently, there are more than 170 known QSOs at $z$$>$5.6 \citep[e.g.][]{Fan2006, Jiang2009, Jiang2016, Willott2010, Banados2014, Banados2015RL, Banados2016, Reed2015, Carnall2015, Venemans2015lowz, Matsuoka2016}, only 12 of which are located at $z$$>$6.5 \citep[][]{Mortlock2011, Venemans2013, Venemans2015highz, Matsuoka2016, Matsuoka2017, Mazzucchelli2017}.
The host galaxies of these very first QSOs are actively forming stars, with prodigious star--formation rates SFR$>$100\,$\msunyr$ \citep[][]{Venemans2012, Venemans2016}, and were able to grow SMBHs with masses exceeding $\mbh$$=$10$^9$\,$\msun$ in less than 800\,Myr \citep[][]{Derosa2014, Venemans2015highz}.
The assembly of such massive SMBHs at early cosmic time requires that they accrete at the Eddington limit (or even super--Eddington) throughout a large fraction of their lifetime \citep[e.g.][]{Yoo2004, Volonteri2005, Volonteri2010, Volonteri2012, Madau2014, Volonteri2015}.
To sustain such a vigorous accretion and intense star formation, the first QSOs need the presence of conspicuous amounts of gas in their surroundings \citep[e.g.][]{Dubois2012, Dimatteo2012}.
Possibly, this gas is aggregated in dense flows able to penetrate into the virial radius of the halo and to funnel gas onto the central SMBH \citep[e.g.,][]{Dimatteo2012, Feng2014}.
If the gas in the host galaxy and in the circum--galactic medium of a QSO is illuminated by the SMBH ionizing radiation and/or from the intense starburst, then it may be observable as an extended ``fuzz'' of fluorescent \lya\ emission \citep{Rees1988, Haiman2001, Alam2002}.

At intermediate redshift ($z$$\sim$2--4), several of these \lya\ nebulae have been reported in the literature, leading to the general consensus that QSOs are frequently (50\%--70\%) embedded in nebulae with sizes of 10--100\,kpc \citep[e.g.][]{Heckman1991a, Heckman1991b, Christensen2006, Hennawi2013, Roche2014, Herenz2015, Arrigoni2016}.
In the last years, gigantic nebulae, with projected sizes $\gtrsim$300\,kpc, have also been detected \citep{Cantalupo2014, Martin2014, Hennawi2015, Borisova2016} suggesting the presence of large amounts of cold gas around intermediate redshift QSOs.
Complementary, the analysis of absorption features in close projected QSO pairs confirms that QSO host galaxies are surrounded (with a covering fraction $\sim$60\% within the virial radius) by cold ($T$$\sim$10$^{4}$\,K), metal enriched ($Z$$\gtrsim$0.1\,$Z_\odot$) gas \citep[][]{Hennawi2006, Hennawi2007, Hennawi2013, Bowen2006, Decarli2009, Prochaska2009, Prochaska2013a, Prochaska2013b, Prochaska2014, Farina2013, Farina2014, Johnson2015, Lau2016, Lau2017}.

Despite the aforementioned achievements, the detection of these structures at $z$$\sim$6 is challenging due to the rapid decrease of the surface brightness (SB) with redshift [SB$\propto$$\left(1+z\right)^{-4}$].
In recent years, large efforts have been made to probe the extended \lya\ emission around $z$$\sim$6 QSOs with contrasting results.
\citet{Decarli2012}, using the Wide Field Camera~3 (WFC3) narrowband filters on the Hubble Space Telescope ({\it HST}), put strong limits on the \lya\ extended emission in the proximity of the highly star forming host galaxies of the QSOs SDSS~J1148+5251 [$z$=6.42, $\llya$$<$2.5$\times$10$^{44}$\,erg\,s$^{-1}$] and SDSS~J1030+0524 [$z$=6.31, $\llya$$<$3.2$\times$10$^{44}$\,erg\,s$^{-1}$].
Conversely, the presence of a \lya\ nebula has been reported in narrowband Suprime-Cam/Subaru images of the QSO CFHQS~J2329$-$0301 at $z$=6.42 \citep{Goto2009} and subsequently confirmed with long--slit spectroscopy with the Echelle Spectrograph and Imager (ESI) and with the DEep Imaging Multi--Object Spectrograph (DEIMOS) spectrographs at the Keck--II telescope \citep{Willott2011, Goto2012}.
Whereas all these observations consistently report the presence of a bright \lya\ halo extending on scales of 15\,kpc in proximity of CFHQS~J2329$-$0301, its luminosity is not well constrained, with values that range from $\llya$$\gtrsim$1.7--7.5$\times$10$^{43}$\,erg\,s$^{-1}$ \citep{Willott2011, Goto2012} up to $\llya$$=$3.6$\times$10$^{44}$\,erg\,s$^{-1}$ (\citealt{Goto2009}, see footnote~10 in \citealt{Decarli2012}).
Recently, \citet{Roche2014} presented long--slit spectroscopic observations with the Optical System for Imaging and low Resolution Integrated Spectroscopy (OSIRIS) mounted on the Gran Telescopio Canarias (GTC) of a sample of QSOs at $z$$>$2, including one at $z$$=$5.95, the radio--loud QSO SDSS~J2228+0110.
While bright, extended \lya\ emission appears ubiquitous in the Roche~et~al. sample at $z$$=$2--3, only a tenuous detection is reported for SDSS~J2228+0110, with a luminosity of $\llya$$\gtrsim$7.8$\times$10$^{42}$\,erg\,s$^{-1}$ extended up to a scale of $\gsim$10\,kpc.
These values should be conservatively considered as lower limits.
Indeed, they were computed by extracting the spectrum of the nebula over a small stripe close to the QSO emission.
Additionally, a proper subtraction of the point--spread--function~(PSF) was hindered by strong sky lines and by the low signal--to--noise ratio (SNR) of the spectrum.

The Integral--field Spectrograph (IFS) Multi Unit Spectroscopic Explorer \citep[MUSE,][]{Bacon2010} on the Very Large Telescope (VLT) is the obvious game changer in this kind of studies.
It produces a spatially resolved spectrum with a spatial sampling of 0\farcs2$\times$0\farcs2 and a nominal spectral resolution ($R$$=$$\lambda$/$\Delta\lambda$) ranging from $R$$=$1750 at 465\,nm to $R$$=$3750 at 930\,nm, allowing to overcome technical limitations of previous spectroscopic and narrowband investigations, such as uncertainties in the systemic redshift of the QSOs, filter and slit losses, and difficulties in performing a proper PSF subtraction.  

In this paper, we present a deep MUSE integration aimed to detect the \lya\ nebular emission around the high redshift QSO \J0305\ ($z$=6.61, with an absolute magnitude at 1450\,\AA\ of $\M1450$$=$$-25.96$$\pm$0.06) discovered by \citet{Venemans2013} using the VISTA Kilo--Degree Infrared Galaxy (VIKING) survey.
Sensitive near--infrared spectroscopy observations obtained with the Folded--port InfraRed Echellette spectrograph (FIRE) mounted on the Magellan Telescope revealed the presence of a SMBH with $\mbh$=(0.95--1.20)$\times$10$^{9}$\,$\msun$ accreting with an Eddington ratio $\lambda_{\rm Edd}$=L$_{\rm Bol}$/L$_{\rm Edd}$=0.68--0.74 \citep{Derosa2014}.
The [\cii]\,158\,$\mu$m emission line [L$_{\rm [\cii]}$=(3.9$\pm$0.2)$\times$10$^{9}$\,$\lsun$, FWHM=255$\pm$12\,km\,s$^{-1}$] and the underlying far--infrared continuum were detected by \citet{Venemans2016} using the Atacama Large Millimeter/submillimeter Array (ALMA).
From these measurements the precise systemic redshift and the SFR of the QSO host galaxy can be inferred.
While QSO redshifts derived from broad emission lines are subject to systematic shifts and large uncertainties \citep[e.g.,][]{Richards2002, Bonning2007, Hewett2010}, especially at $z$$>$6 \citep[][]{Venemans2016, Mazzucchelli2017}, the narrow [\cii] line accurately traces the systemic redshift ($z_{\rm sys}$=6.6145$\pm$0.0001).
In addition, by fitting the FIR continuum spectrum with a modified black body with a spectral index $\beta$=1.6 \citep[after correcting for the impact of the Cosmic Microwave Background, ][]{daCunha2013}, \citet{Venemans2016} obtained a dust temperature of 30\,K and a total far--infrared luminosity of $\ltir$=2.6$\times$10$^{12}$\,$\lsun$.
The investigation of the Spectral Energy Distribution (SED) of high--$z$ QSOs suggests that the dust emission is predominantly powered by star formation especially at $\lambda$$\gsim$100\,$\mu$m rest--frame\citep[e.g.,][; see also \citealt{Valiante2011}]{Leipski2014, Barnett2015}.
This implies\footnote{Assuming a fixed dust temperature of $T_{d}$$=$47\,K \citep[a value commonly observed in high redshift QSO studies; e.g.,][]{Beelen2006}, the estimated FIR luminosity of \J0305\ would be a factor $\sim$2 higher [$\lfir$=(4.0--7.5)$\times$10$^{12}$\,$\lsun$] and the SFR would be in the range 940--1580\,$\msunyr$.
However, current data suggests a lower dust temperature for \J0305\ \citep{Venemans2016}.
In the remainder of the paper, we will thus consider 545\,$\msunyr$ as {\it bona fide} SFR for the QSO's host galaxy.} SFR$_{\rm TIR}$=545\,$\msunyr$.
Aside from PSO~J036+03 \citep{Venemans2015highz, Banados2015}, \J0305\ shows the highest SFR and Eddington ratio among the $z$$>$6.5 QSOs known to date.
It is thus an excellent target to constrain the properties of the gas reservoir that is expected to surround the first QSOs, together with its close environment.

Throughout this paper we assume a concordance cos\-mo\-lo\-gy with H$_0$=70\,km\,s$^{-1}$\,Mpc$^{-1}$, $\Omega_{\rm M}$=0.3, and \mbox{$\Omega_\Lambda$=1-$\Omega_{\rm M}$=0.7}. 
In this cosmology, at $z$=6.6145 the Universe is 0.808\,Gyr old, and an angular scale $\theta$$=$1\arcsec\ corresponds a proper transverse separation of~5.4\,kpc.

\begin{figure*}[tb]
\begin{center}
\includegraphics[width=0.54\textwidth]{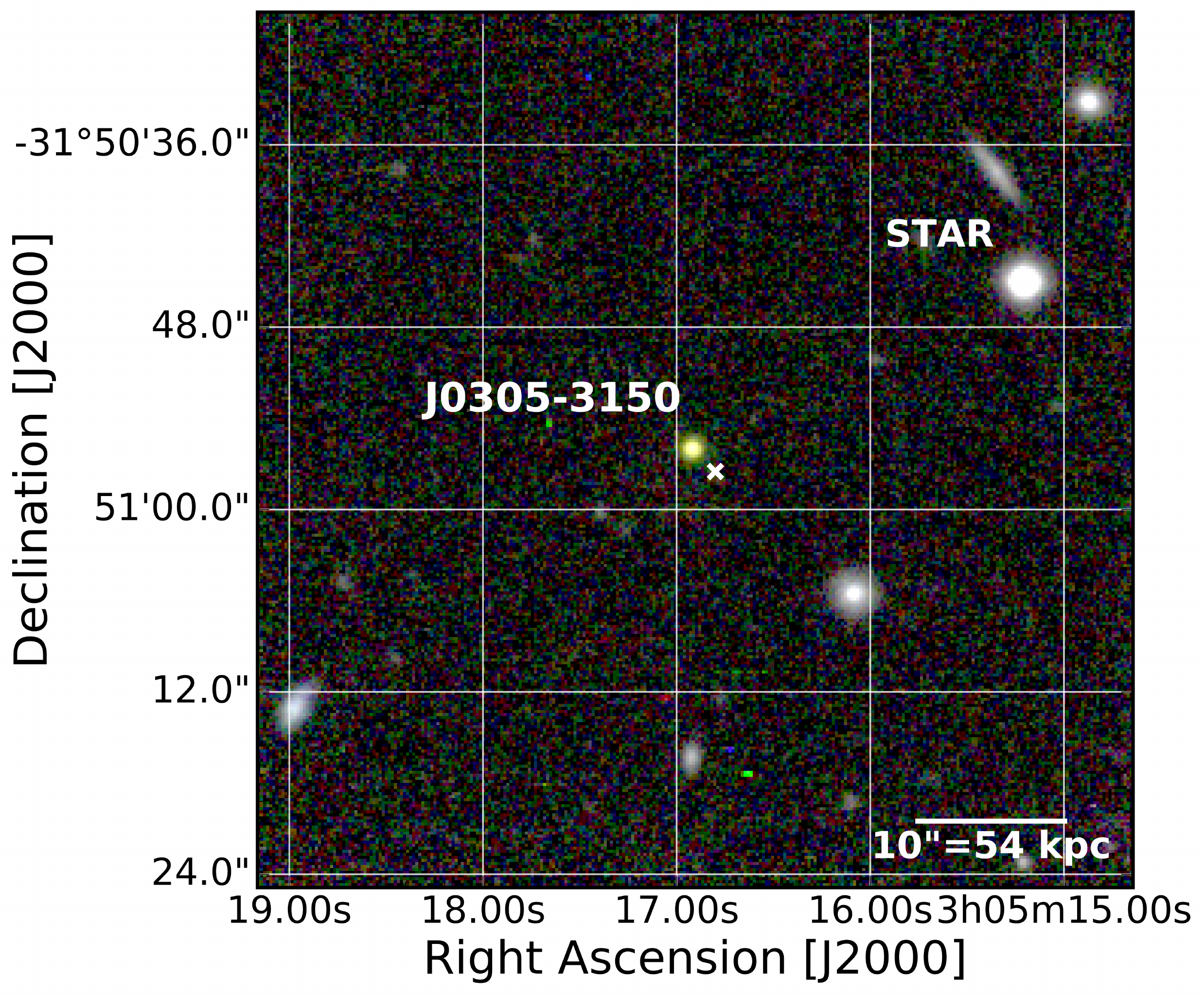}
\includegraphics[width=0.44\textwidth]{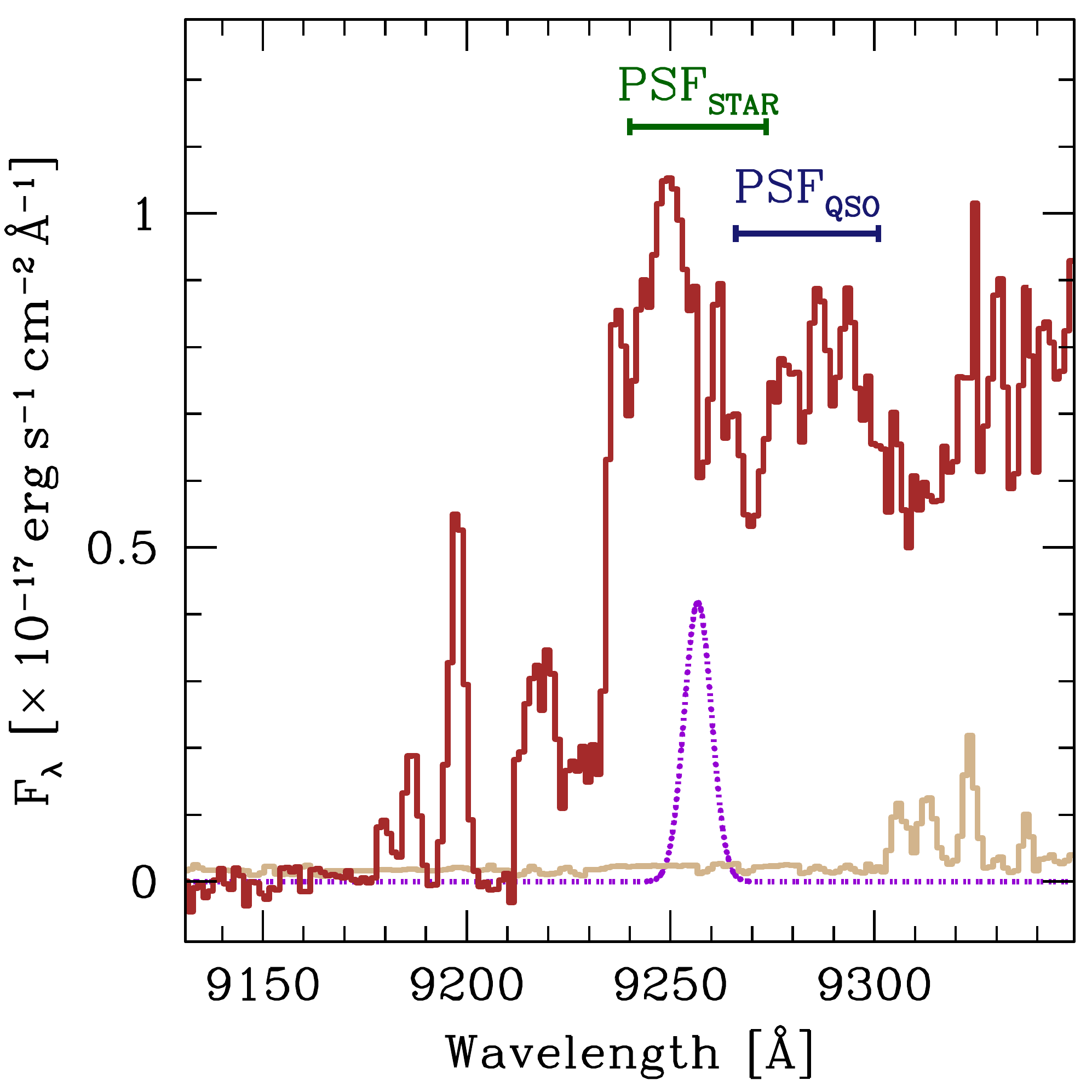}
\caption{
Left Panel --- 
False color RGB image of the field of \J0305\ generated from the MUSE datacube (the wavelength regions used to create the figure are: Red=9262\,\AA\,--\,9292\,\AA, Green=9231\,\AA\,--\,9261\,\AA, and Blue=9200\,\AA\,--\,9230\,\AA).
The QSO and the star used for the PSF subtraction are labeled (see Section~\ref{sec:psf}).
The white cross marks the position of the possible \lya\ emitter identified at $\sim$2\arcsec\ from the QSO (see Section~\ref{sec:lae}).
Right Panel --- 
Zoom--in of the spectrum of \J0305\ extracted over an aperture with 0\farcs74 radius from the MUSE datacube (dark brown solid line), 1--$\sigma$ flux uncertainties are shown in light brown.
The purple dotted line shows a Gaussian emission line at the expected \lya\ location with a FWHM of 255\,km\,s$^{-1}$, as the [CII] line \citep[][, flux normalization is arbitrary]{Venemans2016}, though scattering effects could affect this shape.
The solid blue bar highlights the wavelength range used to construct the empirical PSF$_{\rm QSO}$ model (from 9266\,\AA\ to 9301\,\AA) and the solid green one marks the region used to built PSF$_{\rm STAR}$ (from 9240\,\AA\ to 9273\,\AA, see Section~\ref{sec:psf}). 
}\label{fig:spectrum}
\end{center}
\end{figure*}

\section{OBSERVATIONS AND DATA REDUCTION}\label{sec:observations}

\J0305 was observed with MUSE on 15$^{\rm th}$ December 2014 and 15$^{\rm th}$ January 2015\footnote{Programme ID: 094.B-0893(A), PI: B.~P.~Venemans}.
The total time on source of 2.4\,h was divided in three observing blocks (OB) of 48\,m each, pointing to three different locations within 4\arcsec\ from the QSO.
For each OB, three 960\,s exposures were taken, dithering using random offsets within a 7\farcs5 box.
The Differential Image Motion Monitor (DIMM) seeing during the observations was mostly sub--arcsecond, ranging from 0\farcs7 to 1\farcs1 (with a median of~0\farcs8).

We employed the MUSE Data Reduction Software \citep[version 1.0.1,][]{Weilbacher2012, Weilbacher2014} to perform, on each of the individual exposures, bias subtraction, flat--field, twilight, and illumination corrections, as well as wavelength and flux calibrations (the latter using the standard stars HD49798 and GD71 observed at the beginning of each observing night).
Flat--field correction and sky subtraction of each exposure was improved using the \textsc{CubExtractor} package (S.~Cantalupo et al. in prep.).
Exposures of each OB were sampled to a common grid (0\farcs2$\times$0\farcs2$\times$1.25\,\AA) and then average combined. 
The absolute flux calibration for the resulting datacubes was obtained rescaling the flux of bright sources present in the MUSE FoV to our $i$--band images collected with the ESO Faint Object Spectrograph and Camera~2 (EFOSC2) on the New Technology Telescope \citep[NTT, see][ for details]{Venemans2013}.
Uncertainties on the absolute flux calibration resulting from this procedure are on the order of 5\%.
As a final step, the three datacubes were average combined.
The astrometry solution was refined matching sources in the datacube with the first data release of the VIKING catalogue \citep[][]{Edge2013}.

The MUSE pipeline provides also a datacube containing errors formally propagated throughout the reduction process.
As observed by \citet{Bacon2015}, however, this process does not take into account correlations between neighboring voxels, ending up in underestimating the real noise properties of the datacube. 
To have more realistic uncertainties, for each wavelength slice, the average of the variance delivered by the pipeline was rescaled to match the variance of the background \citep[i.e., after removing the contribution from bright sources; see][ for a similar approach]{Borisova2016}.

In the final datacube, at 9256\,\AA\ (i.e., at the wavelength slice where the \lya\ emission of the QSO is red\-shift\-ed to), the FWHM of the PSF is 0\farcs58, corresponding to 3.1\,kpc at the QSO's redshift.
The 5--$\sigma$ surface brightness limit estimated after collapsing 5~wavelength slices centered at 9256\,\AA\ (i.e., from 9253.5\,AA\ to 9258.5\AA) is SB$_{5\sigma,\lambda}^{1}$=1.9$\times$10$^{-18}$\,erg\,s$^{-1}$\,cm$^{-2}$\,arcsec$^{-2}$ over a 1\,arcsec$^2$ aperture (see Section~\ref{sec:lim}).
The MUSE false color RGB image and the spectrum of \J0305\ extracted over a radius of 3.7\,spaxel (0\farcs74) are showed in Figure~\ref{fig:spectrum}.
Flux errors on the spectrum ($\sigma_{A,\lambda}$) are calculated from the final datacube as: 
\begin{equation}
\sigma_{A,\lambda}=\sqrt{\sum_{i\in  A}\sigma^2_{\lambda,i}}
\end{equation}
where $A$ is the area over which the spectrum is extracted in each wavelength slice ($\lambda$) and $\sigma^2_{\lambda,i}$ is the variance in the corresponding spaxels. 

\section{RECOVERING EXTENDED EMISSION} 

In the following sections we describe the procedure adopted to investigate the presence of an extended \lya\ emission (Section~\ref{sec:psf}) and to estimate the sensitivity reached in the reduced datacube (Section~\ref{sec:lim}).
Finally, in Section~\ref{sec:halo}, we present the results of this analysis.

\subsection{PSF subtraction}\label{sec:psf} 

An accurate PSF subtraction is necessary to recover the faint signal of the diffuse \lya\ halo emerging from the PSF wings of the bright unresolved nuclear component.
To perform this task we created two empirical PSF models: {\it PSF}$_{QSO}$ --- constructed directly from the QSO emission by collapsing regions of the spectrum virtually free from any extended emission\footnote{In principle, the UV continuum light from the QSO host galaxy may contribute to the wings of PSF$_{\rm QSO}$.
However, in $z$$>$5.5 QSOs, this emission is expected to be feeble \citep[e.g.,][]{Mechtley2012}. Given the relatively small wavelength range used to built the PSF model we consider this contribution negligible.} (i.e., away from the \lya\ emission, see~Figure\,\ref{fig:spectrum}); and {\it PSF}$_{STAR}$ --- obtained from the bright star located $\sim$25\arcsec\ North--West from the QSO by summing up its emission over the wavelength range where the extended \lya\ emission is expected to fall (see~Figure\,\ref{fig:spectrum}).
These two PSF models are subject to different systematics, allowing us to check for the reliability of a possible detection of extended emission.
The first model allows us to directly subtract the PSF contribution from the QSO without any spatial shift. 
However, PSF$_{\rm QSO}$ has a relatively low SNR due to the faintness of the source and the small range in wavelength used (starting 325\,km\,s$^{-1}$ away from the QSO systemic redshift, i.e. at 9266\,\AA, up to the wavelength where the presence of strong sky emission lines drastically increase the variance, i.e. 9301\,\AA, see Figure~\ref{fig:spectrum}).
In addition, it may be contaminated by the wings of the possible extended emission if particularly broad and/or redshifted.
PSF$_{\rm STAR}$ instead benefits of a higher SNR and it minimizes PSF changes with wavelength.
On the other hand, it is subject to resampling to centroid the PSF model on the QSO and to the the spatial variation of the PSF.

To subtract the unresolved QSO emission and to recover the \lya\ nebula we adapted the technique used by \citet{Hennawi2013} and \citet{Arrigoni2015} to the 3~dimensional structure of the MUSE data. 
First, at each wavelength slice, the PSF model is rescaled to the QSO's flux estimated in a circle with radius 2\,spaxel (0\farcs4).
The underlying assumption is that the QSO is dominating the emission in this central region.
A bright, centralized nebular component may, however, lead to an overestimate of the QSO's emission and thus to an underestimate of the total flux of the possible extended emission.
Then, we defined the ${\mathlarger \chi}_{\lambda,i}$ datacube:
\begin{equation}\label{eq:chi}
{\mathlarger \chi}_{\lambda,i}=\frac{{\rm DATA}_{\lambda,i}-{\rm MODEL}_{\lambda,i}}{\sigma_{\lambda,i}}
\end{equation}
where the indices $\lambda$ and $i$ indicate the wavelength slice and the 2D spaxel position, respectively; ${\rm DATA}_{\lambda,i}$ is the datacube; ${\rm MODEL}_{\lambda,i}$ is the rescaled PSF model; and $\sigma_{\lambda,i}$ is the square root of the variance datacube.
If our model accurately describes the PSF (and in absence of systematics), at each wavelength slice the distribution of ${\mathlarger \chi}_{\lambda,i}$ values should follows a Gaussian centered in zero with unit variance.
Under this condition, this datacube thus permits to assess the statistical significance of any putative detection.
Note that both PSF$_{\rm QSO}$ and PSF$_{\rm STAR}$ have, by construction, a much higher SNR than the QSO in a single wavelength slice, therefore the contribution of the PSF model to the variance budget is negligible.
We also constructed a smoothed datacube ${\rm SMOOTH}\left[{\mathlarger \chi}_{\lambda,i}\right]$, that is helpful for identifying the possible presence of extended emission:
\begin{equation}\label{eq:schi}
{\rm SMOOTH}\left[{\mathlarger \chi}_{\lambda,i}\right]=\frac{
{\rm CONVOL}\left[{\rm DATA}_{\lambda,i}-{\rm MODEL}_{\lambda,i}\right]}{\sqrt{{\rm CONVOL}^2\left[\sigma^2_{\lambda,i}\right]}}
\end{equation}
where ${\rm CONVOL}$ indicates a convolution in the spatial axis with a 2D Gaussian kernel with $\sigma$=1\,spaxel, while ${\rm CONVOL}^2\left[\sigma^2_{\lambda,i}\right]$ is the convolution of variance datacube ($\sigma^2_{\lambda,i}$) with the square of the same kernel.
In absence of systematics, spaxel values in this smoothed images should still follow a Gaussian distribution, but with smaller variance due to the increased correlation among pixels.
The extended \lya\ nebulae around radio quiet QSO typically show quiescent kinematics with FWHM$\lesssim$600\,km\,s$^{-1}$ \citep[][]{Arrigoni2015giant, Borisova2016}.
We therefore repeat the same procedure binning over 3, 5, 10, and~15 wavelength slices in order to maximize the SNR of emission with FWHM of $\sim$120, 200, 400, and 600\,km\,s$^{-1}$.
The resulting ${\mathlarger \chi}_{\lambda,i}$ and ${\rm SMOOTH}\left[{\mathlarger \chi}_{\lambda,i}\right]$ datacubes were then visually inspected to search for the presence of any extended emission.

\begin{figure}[tb]
\begin{center}
\includegraphics[width=0.46\textwidth]{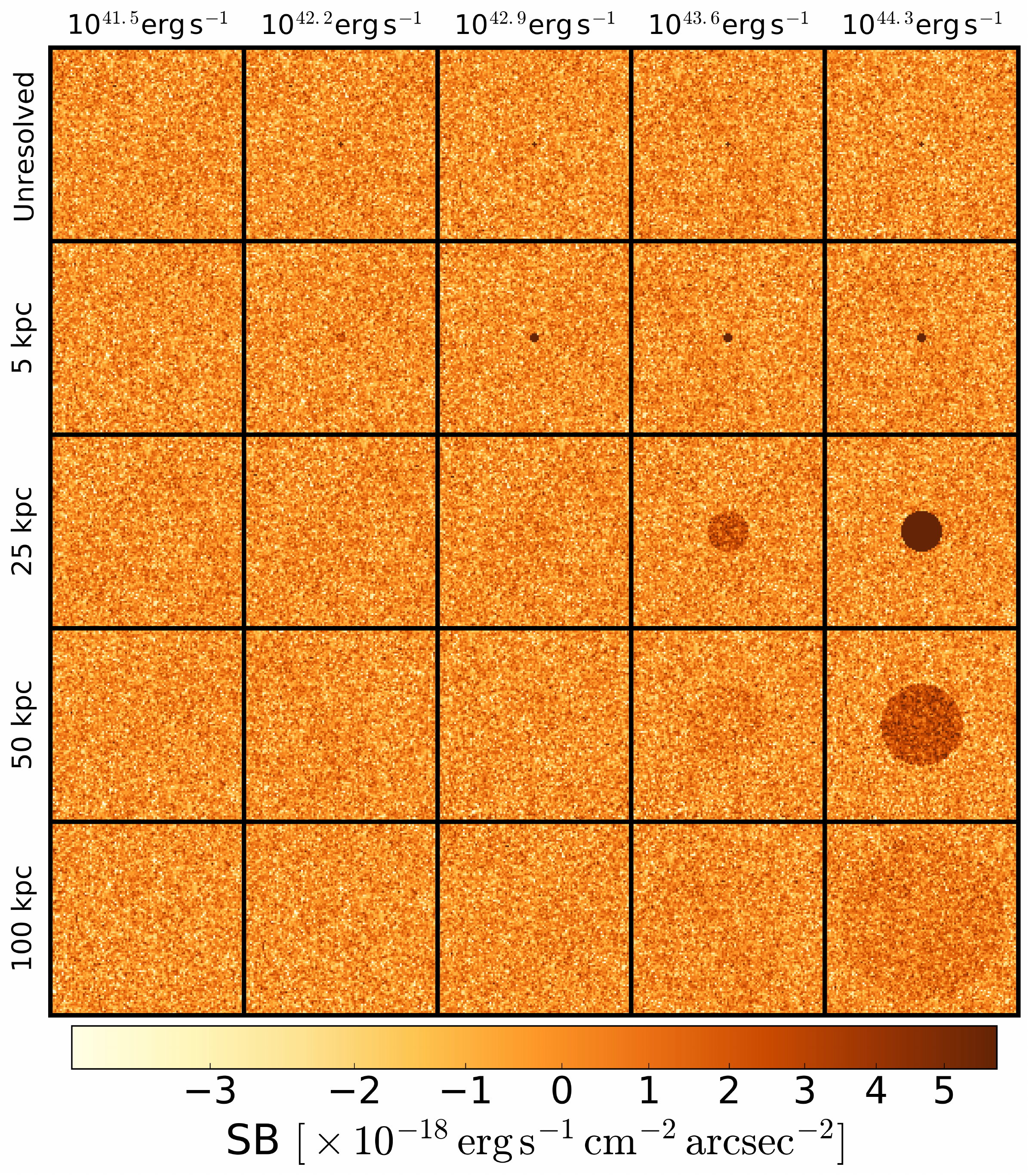}
\caption{
Illustration of how synthetic sources, created using the procedure described in Section~\ref{sec:lim}, would appear in a pseudo narrowband image obtained by binning the $\left[{\rm DATA}_{\lambda,i}-{\rm MODEL}_{\lambda,i}\right]$ datacube over three wavelength slices (i.e. $\sim$120\,km\,s$^{-1}$) around the expected position of the \lya\ emission.
The nominal 5--$\sigma$ surface brightness limit reached in this pseudo narrowband image is SB$_{5\sigma,\lambda}^1$=1.3$\times$10$^{-18}$\,erg\,s$^{-1}$\,cm$^{-2}$\,arcsec$^{-2}$ over a 1\,arcsec$^2$ aperture.
At the center of each box is plotted a source with fix FWHM$_{\rm mock}$=120\,km\,s$^{-1}$, total luminosity ranging from L$_{\rm mock}$=10$^{41.5}$\,erg\,s$^{-1}$ to 10$^{44.3}$\,erg\,s$^{-1}$ (increasing from the left to the right of the x--axis), and diameter going from d$_{\rm mock}$=3\,kpc (i.e., unresolved) to 100\,kpc (increasing from top to bottom of the y--axis).
Note that, assuming a Gaussian shape for the line emission, only roughly half of the total flux falls in the pseudo narrowband image shown here.
Each box has a size of 22\arcsec$\times$22\arcsec.
}\label{fig:lim1}
\end{center}
\end{figure}

\begin{figure}[tb]
\begin{center}
\includegraphics[width=0.46\textwidth]{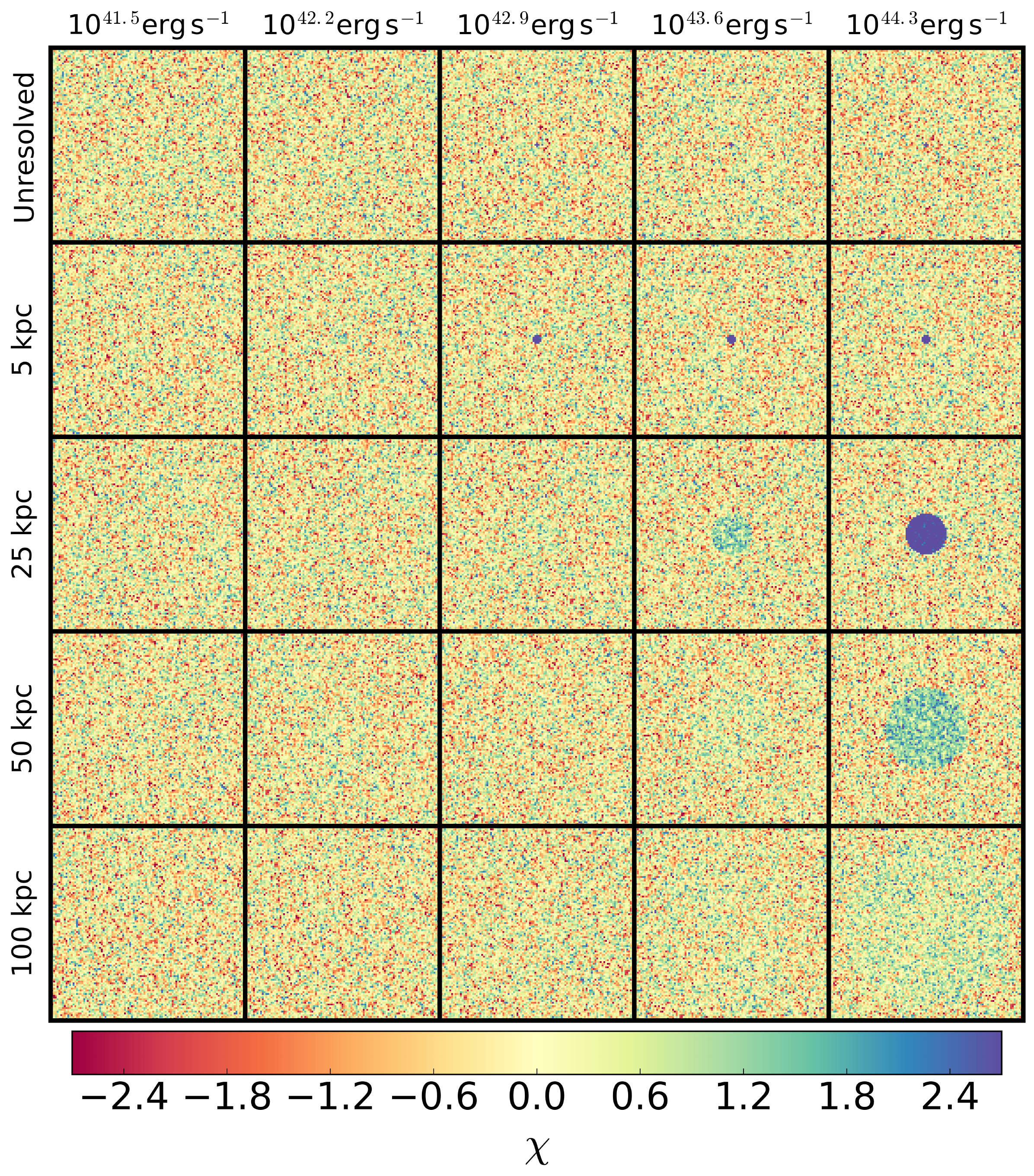}
\newline
\includegraphics[width=0.46\textwidth]{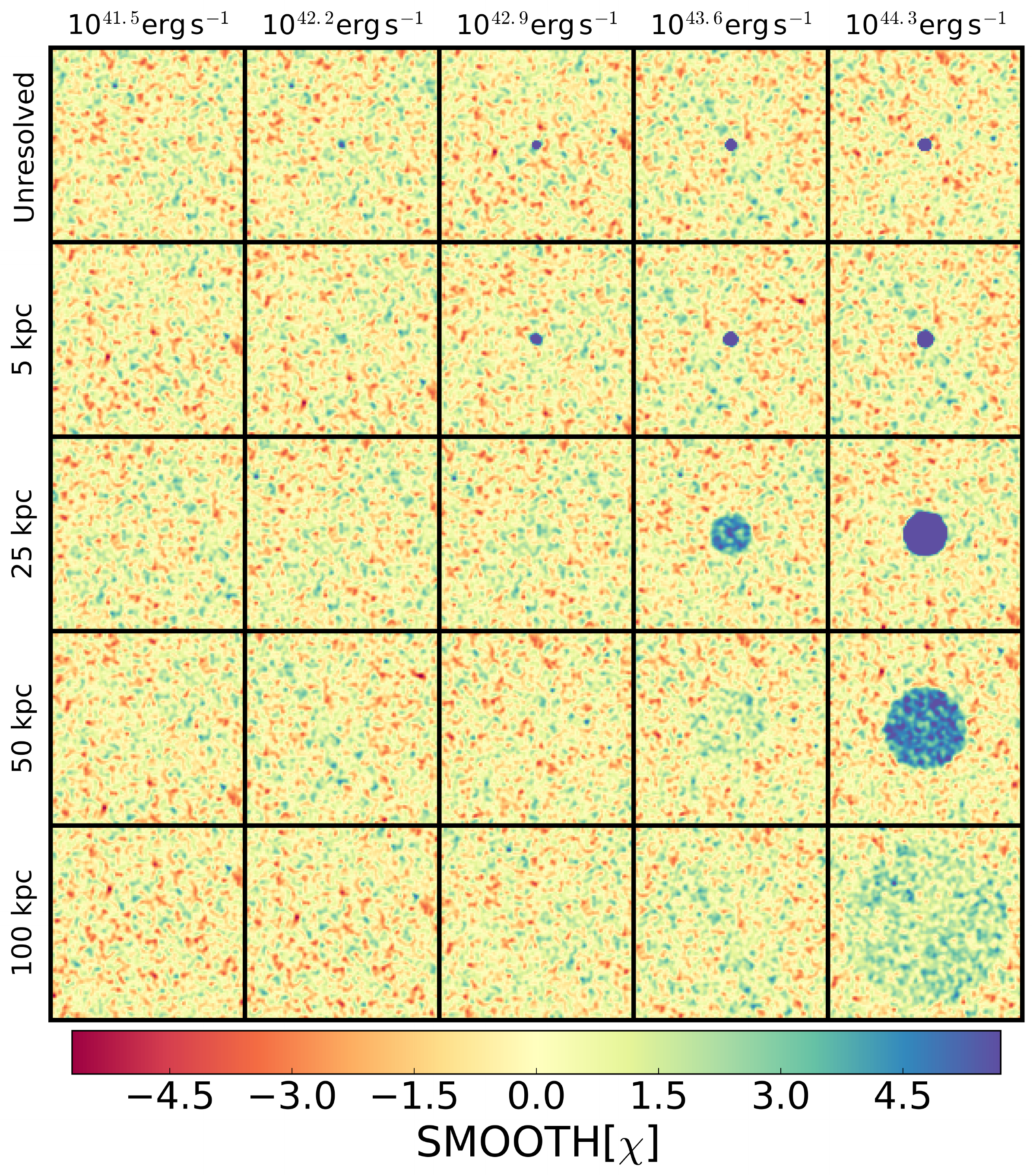}
\caption{
${\mathlarger \chi}_{i}$ (see Equation~\ref{eq:chi}, Top Panel) and ${\rm SMOOTH}\left[{\mathlarger \chi}_{i}\right]$ (see Equation~\ref{eq:schi}, Bottom Panel) pseudo narrowband images of the synthetic sources showed in Figure~\ref{fig:lim1}.
In absence of systematics, ${\mathlarger \chi}_{i}$ displays the statistical significance of the emission in each spaxel.
The smooth process introduce correlation among neighbor spaxels.
This enhance the coherent signal coming from close positive spaxels, allowing us to increase our ability to detect faint extended sources (note the different scale of the colorbars).
At a fixed luminosity, this process is more sensitive to compact objects rather then diffuse emissions.
The nominal 5--$\sigma$ detection limits are ${\rm L}_{5\sigma}$=10$^{[41.9,\,42.1,\,42.8,\,43.1,\,43.4]}$\,erg\,s$^{-1}$ for sources with d$_{\rm mock}$=[3, 5, 25, 50, 100]\,kpc, respectively.
The size of each box is the same of Figure~\ref{fig:lim1}.
}\label{fig:lim2}
\end{center}
\end{figure}

\begin{figure}[ht]
\begin{center}
\includegraphics[width=0.48\textwidth]{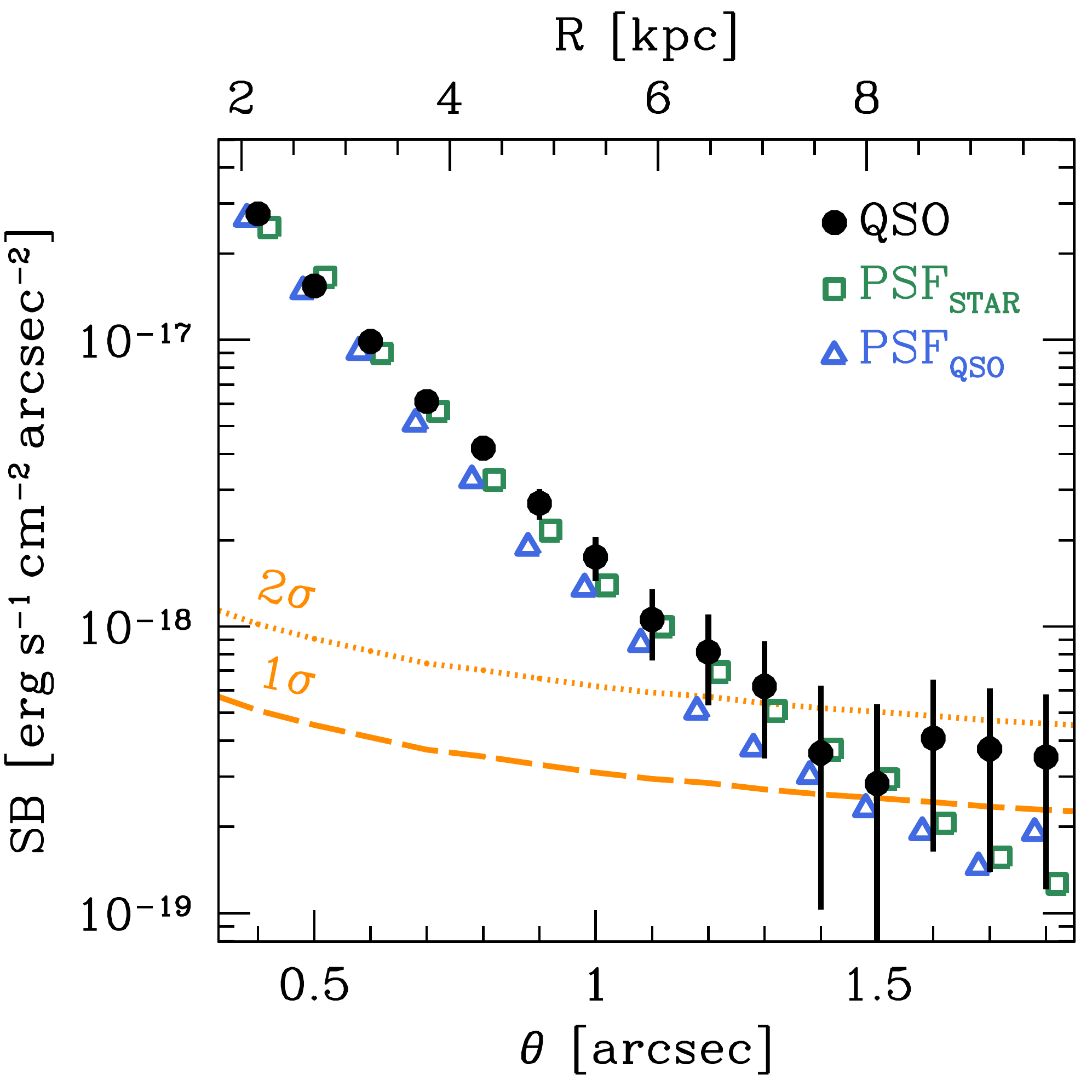}
\caption{
Radial profile of the QSO emission extracted within annuli evenly spaced of 0\farcs2 in the pseudo narrowband image created collapsing the datacube over 3~wavelength slices between 9238.5\,\AA\ and 9242.0\,\AA\ (i.e., where the QSO emission peaks and contamination from the possible extended emission is expected to be absent; black dots).
For comparison the rescaled PSF$_{\rm STAR}$ (green squares) and PSF$_{\rm QSO}$ (blue triangles) are also shown (points are artificially shifted on the x--axis to avoid superposition).
The two PSF models appear in good agreement with the QSO profile.
The region at angular separation $\theta$$<$0\farcs4 is used to normalize the PSF models (see Section~\ref{sec:psf}) and therefore is not plotted here.
The nominal 5--$\sigma$ surface brightness limit reached in this image is SB$_{5\sigma,\lambda}^1$=1.2$\times$10$^{-18}$\,erg\,s$^{-1}$\,cm$^{-2}$\,arcsec$^{-2}$ over a 1\,arcsec$^2$ aperture.
The orange dashed and dotted lines mark the corresponding 1-- and 2--$\sigma$ limits, respectively.
The wings of the QSO emission become consistent with the noise at separation larger then $\gtrsim$1\farcs5 (at 1--$\sigma$).
Our ability to detect extended emission at these scales is thus not influenced by the PSF subtraction procedure.
}\label{fig:radial}
\end{center}
\end{figure}

\subsection{Detection Limits}\label{sec:lim}

The detection limit for a \lya\ nebular emission in the MUSE datacube depends on its FWHM and physical size: lower surface brightness levels can be reached averaging in space and/or in velocity.
Under the (erroneous) assumption that spaxels are independent, the theoretical detection limits for an extended source in a single wavelength slice is given by SB$_{1\sigma,\lambda}^{A}$=SB$_{1\sigma,\lambda}$/$\sqrt{\#_A}$, where SB$_{1\sigma,\lambda}$ is the 1--$\sigma$ surface brightness detection limit per 0\farcs2$\times$0\farcs2 spaxel at the wavelength slice $\lambda$, and $\#_A$ is the number of spaxels in the isophotal area of the source $A$.
The corresponding limit on the total flux (and thus on the luminosity) can be written as: F$_{1\sigma,\lambda}^{A}$=SB$_{1\sigma,\lambda}$\,$\sqrt{\#_A}$\,${\rm PS}^2$, where ${\rm PS}$ is the pixel scale of MUSE: ${\rm PS}$=0\farcs2\,spaxel$^{-1}$.
Binning the MUSE datacube over 3, 5, 10, and~15 wavelength slices centered at 9256\,\AA\ (i.e. at the expected position of the \lya\ emission), the formal 5--$\sigma$ surface brightness detection limits calculated over an aperture of 1\,arcsec$^2$ are: SB$_{5\sigma,\lambda}^{1}$=[1.3, 1.9, 3.4, 5.0]$\times$10$^{-18}$\,erg\,s$^{-1}$\,cm$^{-2}$\,arcsec$^{-2}$, respectively.  

The noise properties of the MUSE datacube are, however, not uniform.
In addition, cross--talk between voxels and systematics introduced during data reduction and PSF subtraction will alter these theoretical detection limits.
We tested the reliability of the calculated surface brightness limits introducing a set of synthetic sources in the PSF--subtracted datacubes and visually estimated the level of a convincing detection.
For this purpose we focused on the wavelength region where a \lya\ line redshifted to $z_{\rm sys}$ would fall (i.e. at 9256\,\AA) and we binned the datacube over the same wavelength slices used for the sources detection (see Section\,\ref{sec:psf}).
We randomly placed mock circular sources (including Poisson noise) with a top--hat surface brightness in different location of each pseudo narrowband images.
These mock sources have total integrated luminosities L$_{\rm mock}$=10$^{[41.5,\,42.2,\,42.9,\,43.6,\,44.3]}$\,erg\,s$^{-1}$, diameters d$_{\rm mock}$=[3, 5, 25, 50, 100]\,kpc, and, in the wavelength space, a Gaussian distribution with FWHM$_{\rm mock}$=[80, 120, 200, 400, 600]\,km\,s$^{-1}$ (where 80\,km\,s$^{-1}$ is, roughly, the nominal resolution limit and the other values match the binning considered in the PSF subtraction process, see Section~\ref{sec:psf}).
A diameter of d$_{\rm mock}$=3\,kpc corresponds to the seeing measured in the datacube and is hence unresolved. 

As an illustrative example we show in Figure~\ref{fig:lim1} how synthetic sources with FWHM$_{\rm mock}$=120\,km\,s$^{-1}$ and different d$_{\rm mock}$ and L$_{\rm mock}$ would appear in a pseudo narrowband image obtained collapsing three wavelength slices around the expected position of the \lya\ line (i.e. from 9254.75\,\AA\ to 9257.25\,\AA).
The corresponding ${\mathlarger \chi}_{i}$ and ${\rm SMOOTH}\left[{\mathlarger \chi}_{i}\right]$ images (Figure~\ref{fig:lim2}) show that we should be able to visually detect these sources down to $\llya$$\sim$10$^{42.0}$\,erg\,s$^{-1}$ if unresolved and down to $\llya$$\sim$10$^{44.0}$\,erg\,s$^{-1}$ if the emission is more extended. 

If sources fall on the top of the QSO emission, the PSF subtraction process will hinder the achievement of the theoretical detection limits.
However, the QSO is relatively faint in the datacube and PSF wings quickly drop below the 2--$\sigma$ surface brightness limit. 
For instance, in the pseudo narrowband obtained collapsing the datacube between 9238.5\,\AA\ and 9242.0\,\AA\ the QSO radial profile becomes consistent with zero (at 2--$\sigma$) at a separation of $\sim$1\farcs2 (i.e., $\sim$6.5\,kpc at $z_{\rm sys}$, see Figure~\ref{fig:radial}).
Contamination due to imperfect PSF subtraction are therefore expected to impact only sub--arcseconds separations.

\begin{figure*}[tb]
\begin{center}
\includegraphics[width=0.305\textwidth]{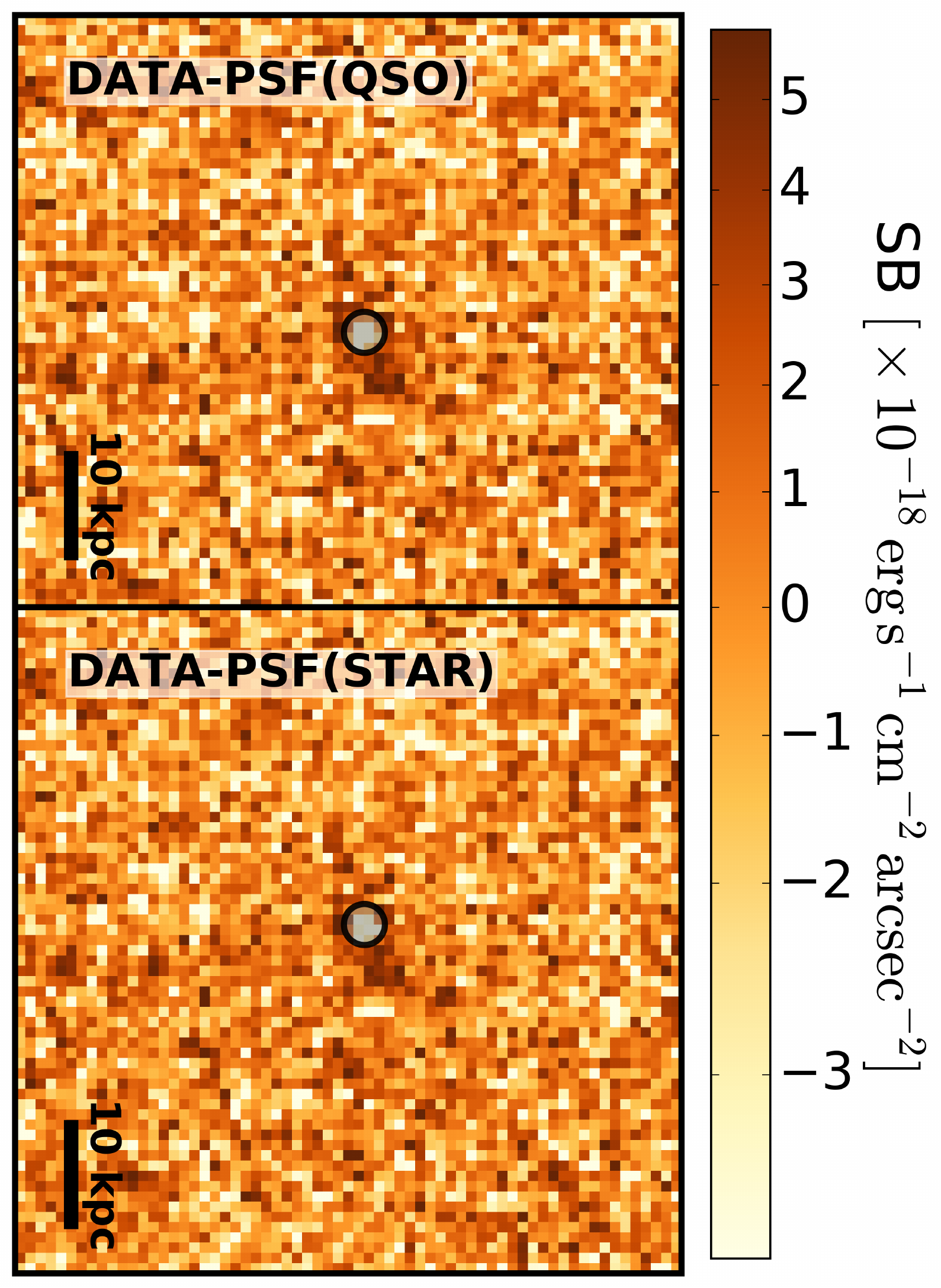}
\includegraphics[width=0.32\textwidth]{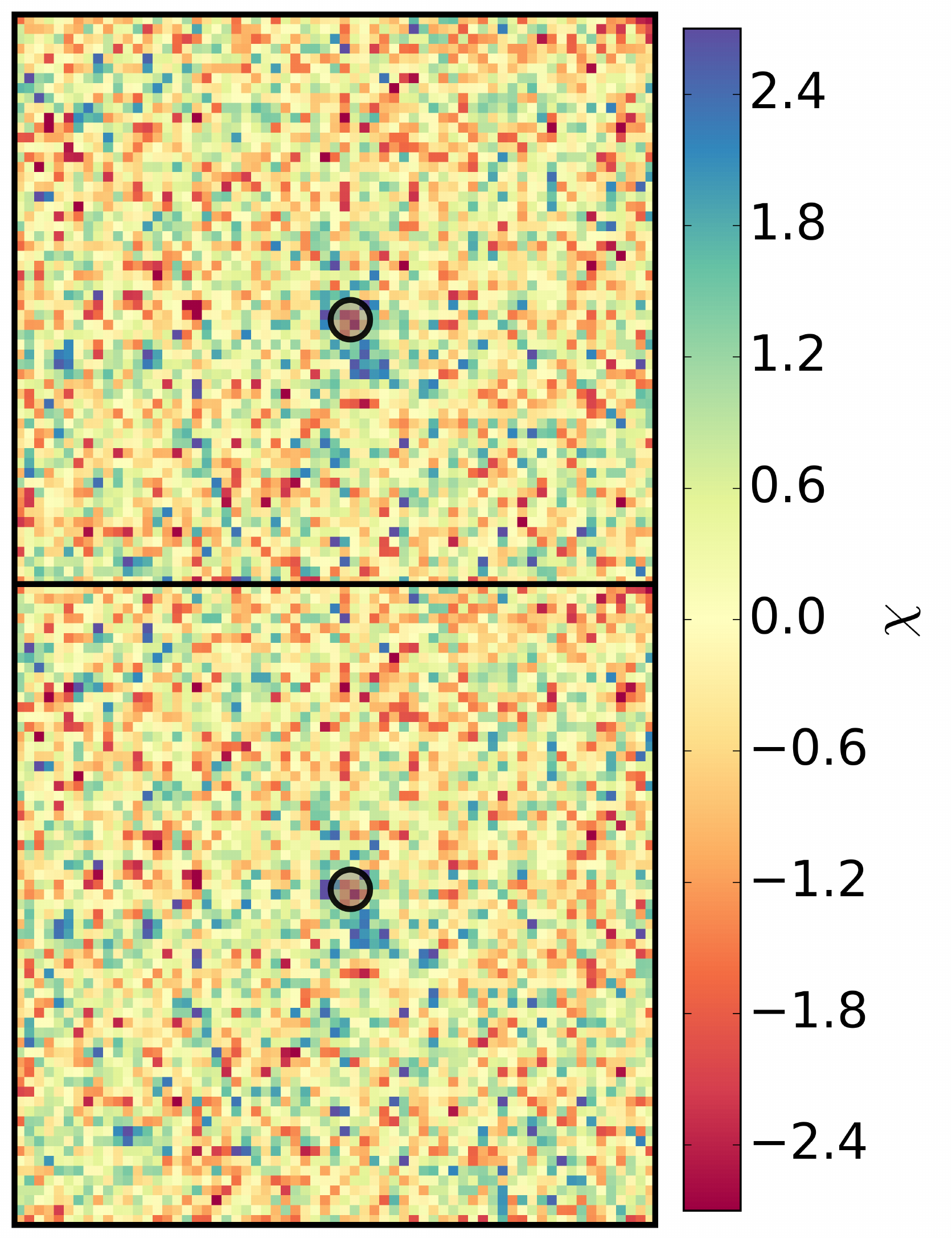}
\includegraphics[width=0.32\textwidth]{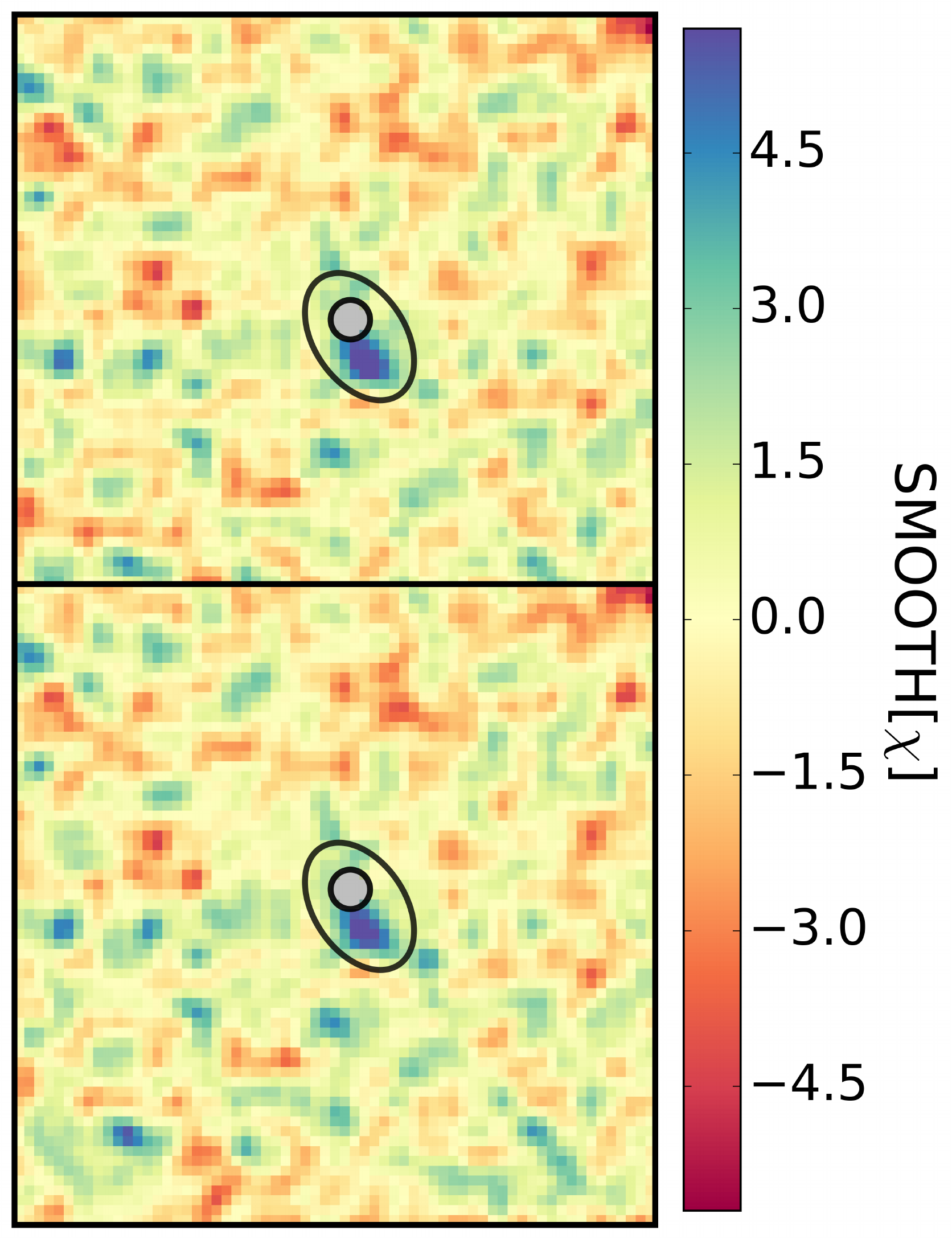}
\caption{
Result of the PSF subtraction procedure in the wavelength range where the presence of an extended \lya\ halo is tentatively detected (from 9259.75\,\AA\ to 9262.25\,\AA, i.e. over 3 wavelength slices, see Figure~\ref{fig:halospec}).
The different panels show (from left to right) \mbox{${\rm DATA}_{i}-{\rm MODEL}_{i}$}, ${\mathlarger \chi}_{i}$, and ${\rm SMOOTH}\left[{\mathlarger \chi}_{i}\right]$ pseudo narrowband images obtained using PSF$_{\rm QSO}$ (top row) and PSF$_{\rm STAR}$ (bottom row).
The size of the boxes is 13\arcsec$\times$13\arcsec\ (70\,kpc$\times$70\,kpc at the QSO's redshift).
Images are oriented to have north on the top and east on the left.
The color scale used here is the same as in Figure~\ref{fig:lim1} and~\ref{fig:lim2}.
The area used to normalize the PSF models to the QSO is marked with a black circle.
This region was masked out to produce the ${\rm SMOOTH}\left[{\mathlarger \chi}_{i}\right]$ images.
The elliptical aperture used to derive the photometry of the extended emission is shown in the rightmost panels.
The 5--$\sigma$ nominal surface brightness limit reached in this pseudo narrowband image is SB$_{5\sigma}^1$=2.2$\times$10$^{-18}$\,erg\,s$^{-1}$\,cm$^{-2}$\,arcsec$^{-2}$ over a 1\,arcsec$^2$ aperture.
A significative excess of residuals is present towards the South--West considering both PSF models.
We notice that our procedure slightly oversubtracts the QSO emission.
The total flux coming from the putative halo could thus be underestimated.
}\label{fig:halo}
\end{center}
\end{figure*}

\begin{figure}[tb]
\begin{center}
\includegraphics[width=0.46\textwidth]{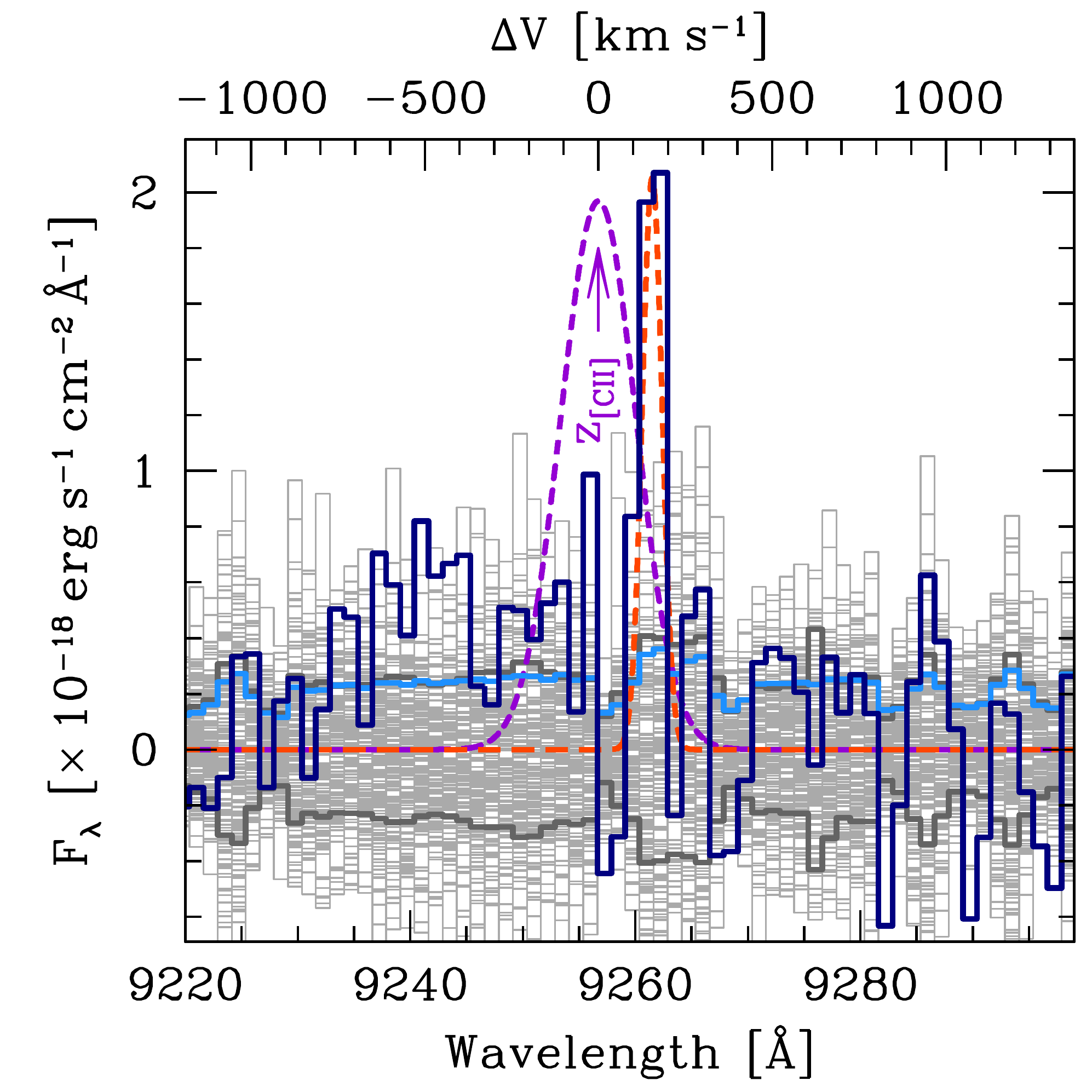}
\caption{
Spectrum of extended \lya\ emission extracted over an elliptical aperture with semi--minor axis of 0\farcs9 and semi--major axis of 1\farcs4 in the ${\rm DATA}_{\lambda,i}-{\rm MODEL}_{\lambda,i}$ datacube (blue solid line), 1--$\sigma$ flux uncertainties are shown in light blue.
A circle with radius 0\farcs4 centered at the QSO position is removed before extracting the spectrum (see Figure~\ref{fig:halo}).
The top axis ($\Delta$V) indicates the velocity shift with respect to the QSO's systemic redshift.
As in Figure~\ref{fig:spectrum}, the purple dashed gaussian shows the expected position and width of the \lya\ line if centered at $z_{\rm sys}$ and with a FWHM equal to the [\cii] line.
The possible \lya\ halo appears as a bright spike at $\lambda$=9261.5\,\AA. 
A gaussian model of the line is shown in red.
Light gray histograms are spectra randomly extracted within a radius of 25\arcsec\ from the QSO's position (see Section~\ref{sec:halo} for details).
The 1--$\sigma$ dispersion of these spectra in each voxel is highlighted in dark gray.
The excess of flux at $\sim$9240\,\AA\ appears to be related to correlations present in the MUSE data and/or to non optimal sky--subtraction and thus non--significative.
}\label{fig:halospec}
\end{center}
\end{figure}

\begin{figure}[tb]
\begin{center}
\includegraphics[width=0.46\textwidth]{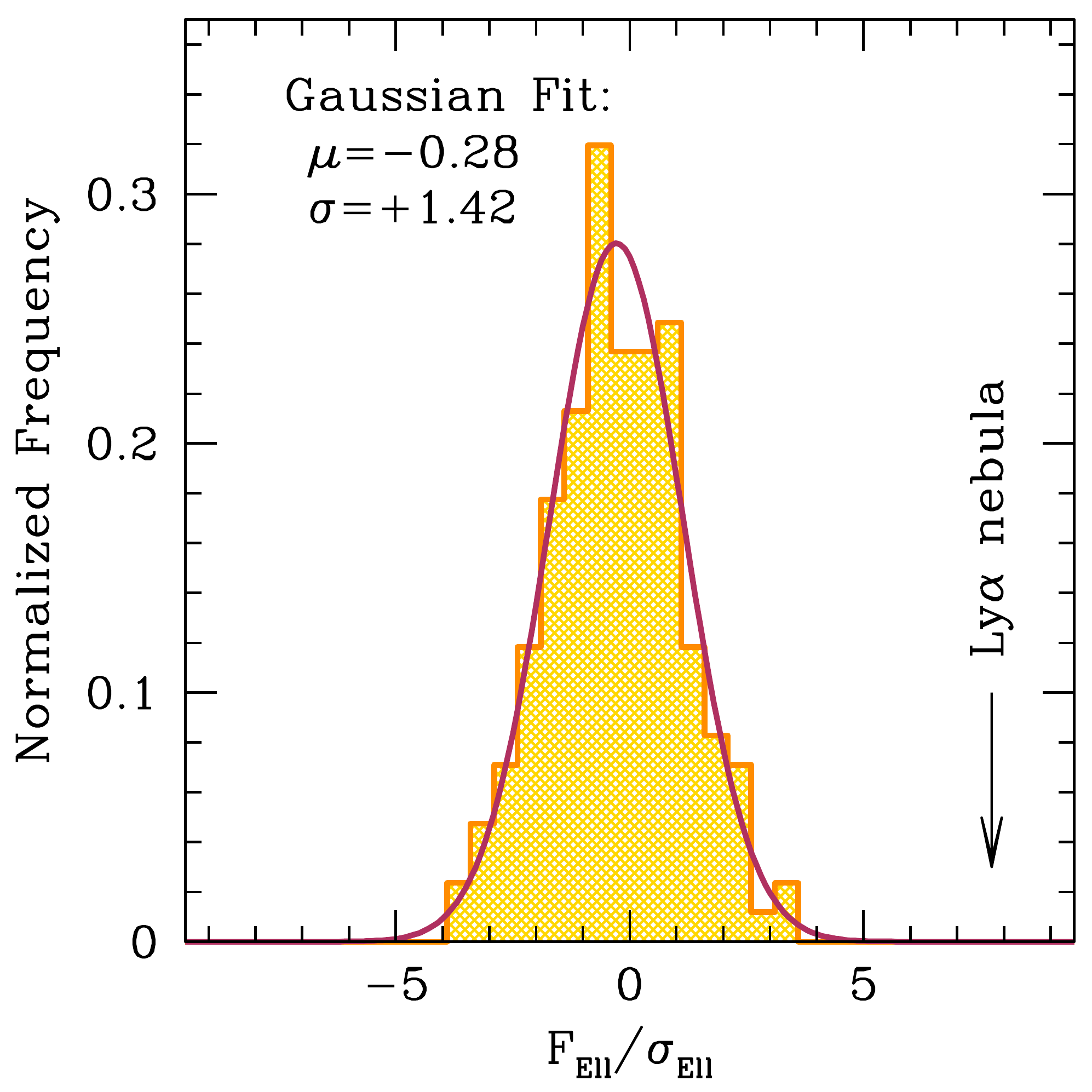}
\caption{
Analysis of the significance of the possible \lya\ nebula detected in the MUSE datacube.
The orange histogram is the distribution of flux over noise of elliptical apertures, with the same extent of the one used to extract the extended emission (see Section~\ref{sec:halo}), randomly placed in the pseudo narrowband image obtained collapsing \mbox{${\rm DATA}_{i}-{\rm MODEL}_{i}$} from 9259.75\,\AA\ to 9262.25\,\AA.
The arrow marks the position of the tentatively detected halo.
The Gaussian fit of the distribution (purple line) has an average of \mbox{$\mu$=-0.28} and a sigma of $\sigma$=1.42.
These shifts from the expected values ($\mu$=0, $\sigma$=1) reflect systematics in the final datacube due, for instance, to poor sky--subtraction and to correlation among voxels.
Taking into account this distribution, the significance of the detection is therefore 5.1--$\sigma$ for PSF$_{\rm QSO}$ and 4.7--$\sigma$ for PSF$_{\rm STAR}$.
}\label{fig:halosyst}
\end{center}
\end{figure}

\subsection{Tentative Detection of an Extended Emission}\label{sec:halo}

Figure~\ref{fig:halo} shows the result of the PSF subtraction procedure described above.
A \lya\ extended emission with a size of $\sim$1\farcs6 ($\sim$9\,kpc) towards the South--West of \J0305\ is tentatively detected in the \mbox{${\rm DATA}_{i}-{\rm MODEL}_{i}$} pseudo narrowband image obtained collapsing the 3~wavelength slices from 9259.75\,\AA\ to 9262.25\,\AA.
After removing a circle of 0\farcs4 radius centered on the QSO's position (to avoid possible contamination due to imperfect PSF subtraction) we measured the flux integrated over an elliptical aperture, with semi--minor and semi--major axis of 0\farcs9 and 1\farcs4.
We obtained $\flya$=(6.1$\pm$0.8)$\times$10$^{-18}$\,erg\,s$^{-1}$\,cm$^{-2}$ [$\flya$=(5.6$\pm$0.8)$\times$10$^{-18}$\,erg\,s$^{-1}$\,cm$^{-2}$] using PSF$_{\rm QSO}$ (PSF$_{\rm STAR}$) as our model, yielding a \mbox{7.6--$\sigma$} \mbox{(7.0--$\sigma$)} detection.
The inferred lu\-mi\-no\-si\-ty of the extended emission is $\llya$=(3.0$\pm$0.4)$\times$10$^{42}$\,erg\,s$^{-1}$ (using PSF$_{\rm QSO}$), more than one order of magnitude fainter with respect to the characteristic luminosity of LAEs at $z$$\sim$6.6 \citep[e.g.][]{Hu2010, Matthee2015}.

In order to assess the reliability of this detection, we empirically estimate the effects of systematics in the pseudo narrowband image obtained collapsing the \mbox{${\rm DATA}_{\lambda,i}-{\rm MODEL}_{\lambda,i}$} datacube over the wavelength range 9259.75\,\AA\ to 9262.25\,\AA\ \citep[see][ for a similar test performed on narrowband images]{Arrigoni2015}.
From this image, we extracted fluxes and variances for a set of elliptical apertures randomly picking background locations (i.e. avoiding bright sources) in an annulus with internal and external radius of 5\arcsec\ and 25\arcsec\ from the QSO, respectively.
In absence of systematics, the flux over noise ratio of these apertures should follow a Gaussian distribution centered in zero and with $\sigma$=1.  
Figure~\ref{fig:halosyst} shows that the actual distribution is nearly Gaussian with an offset of -0.28 and a sigma 1.42$\times$ broader.
These deviations could be due to poor sky--subtraction and/or to 3D correlations present in the MUSE datacube.
Taking into account systematics, the significance of our detection is reduced to 5.3--$\sigma$ (4.9--$\sigma$) using PSF$_{\rm QSO}$ (PSF$_{\rm STAR}$) as PSF model.

The spectrum of this possible nebula extracted from the \mbox{${\rm DATA}_{\lambda,i}-{\rm MODEL}_{\lambda,i}$} datacube over the same elliptical aperture considered above is shown in Figure~\ref{fig:halospec}.
The possible \lya\ halo appears as the strongest feature present within $\pm$1000\,km\,s$^{-1}$ from the QSO's systemic redshift.
This narrow emission line peaks at $\lambda$=9261.5\,\AA\ and has a FWHM=65\,km\,s$^{-1}$.
The fitted width is slightly smaller than the nominal resolution limit of MUSE at these wavelengths, but consistent given the low signal--to--noise per spectral bin of the line.
Under the assumption that we are probing the \lya\ emission, the line appears redshifted by 155\,km\,s$^{-1}$ with respect to the systemic redshift traced by the [\cii] line.
This is smaller than the 445\,km\,s$^{-1}$ shift observed between the \ha\ emission (originated from HII regions in the galaxy) and the \lya\ line in low redshift UV--selected galaxies \citep[e.g.][]{Steidel2010}, but in agreement with the 175\,km\,s$^{-1}$ offset seen in strong \lya\ emitters (LAE) at $z$$\sim$2 \citep[e.g.][]{Hashimoto2013}.
A similar shift is also observed at higher redshift.
For instance, \citet{Pentericci2016} observed shifts of 100--200\,km\,s$^{-1}$ between the \lya\ and [\cii] emission line in four galaxies at redshifts between 6.6 and 7.2.
Likewise, the \lya\ emission of the $z$$\sim$6 starforming galaxy A383-5.2 appears to be shifted by 120\,km\,s$^{-1}$ with respect to the systemic redshift probed via the \ciii]\,$\lambda$1909 metal line \citep[][]{Stark2015}.

\begin{figure*}[tb]
\begin{center}
\includegraphics[width=0.32\textwidth]{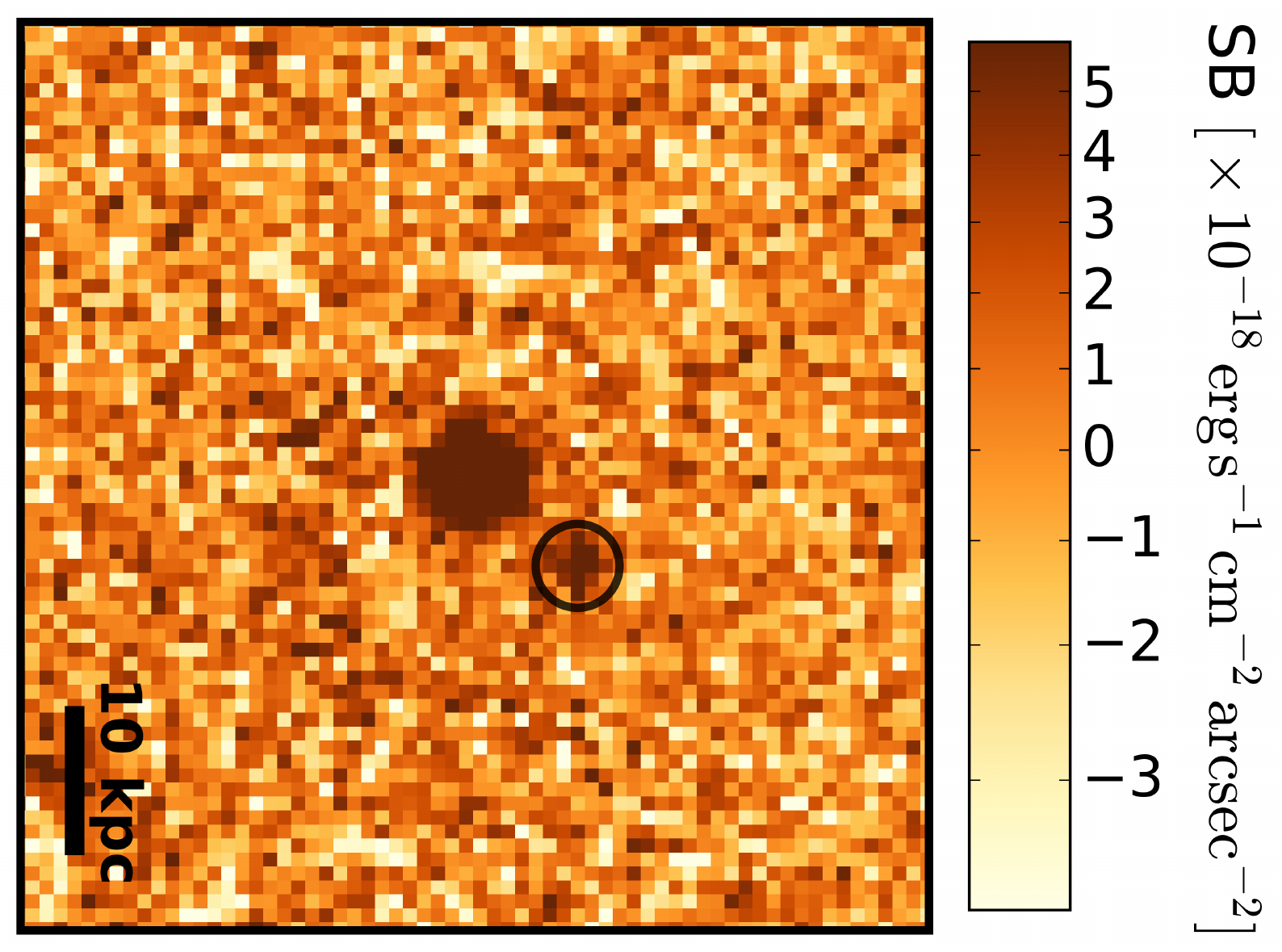}
\includegraphics[width=0.33\textwidth]{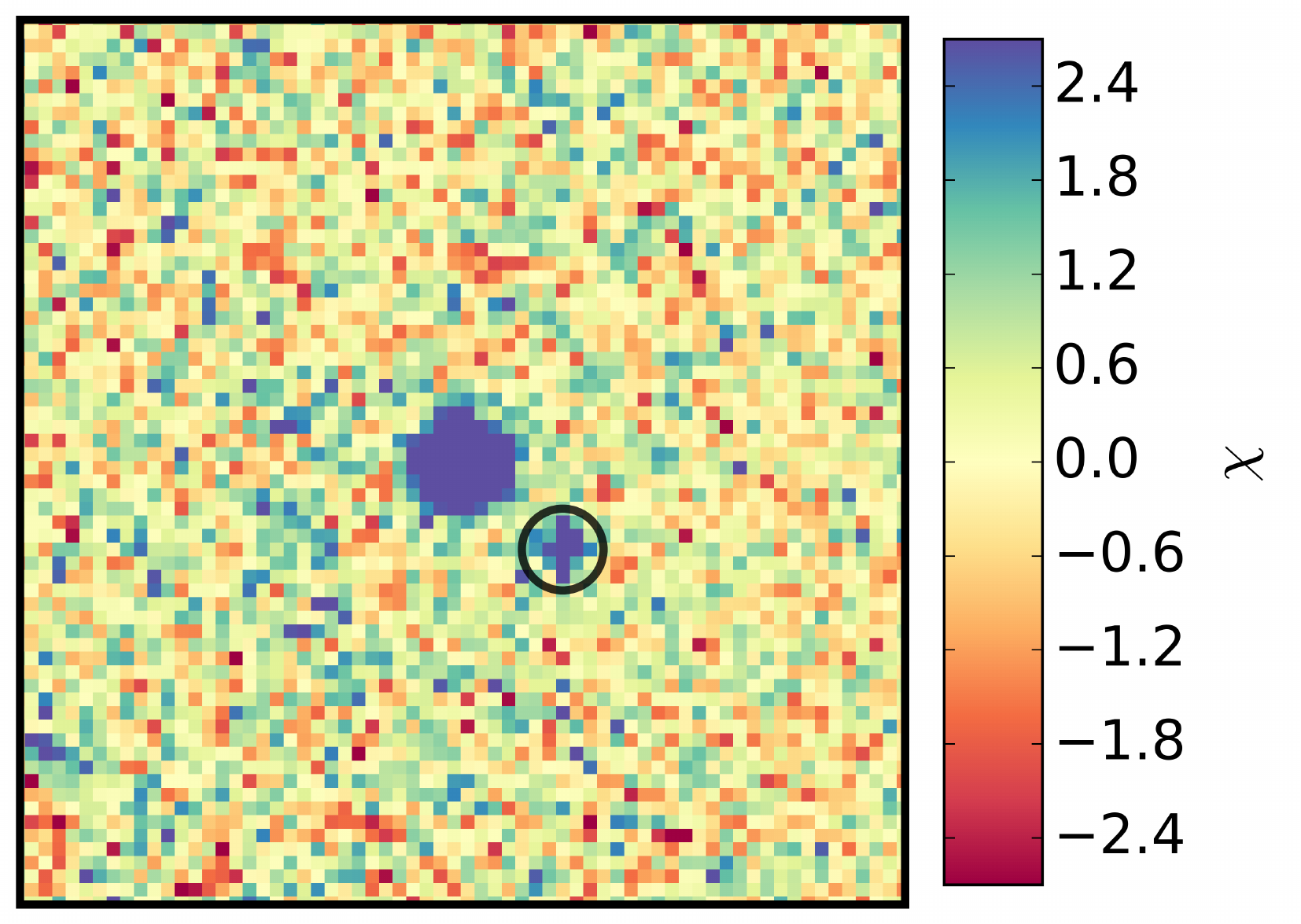}
\includegraphics[width=0.33\textwidth]{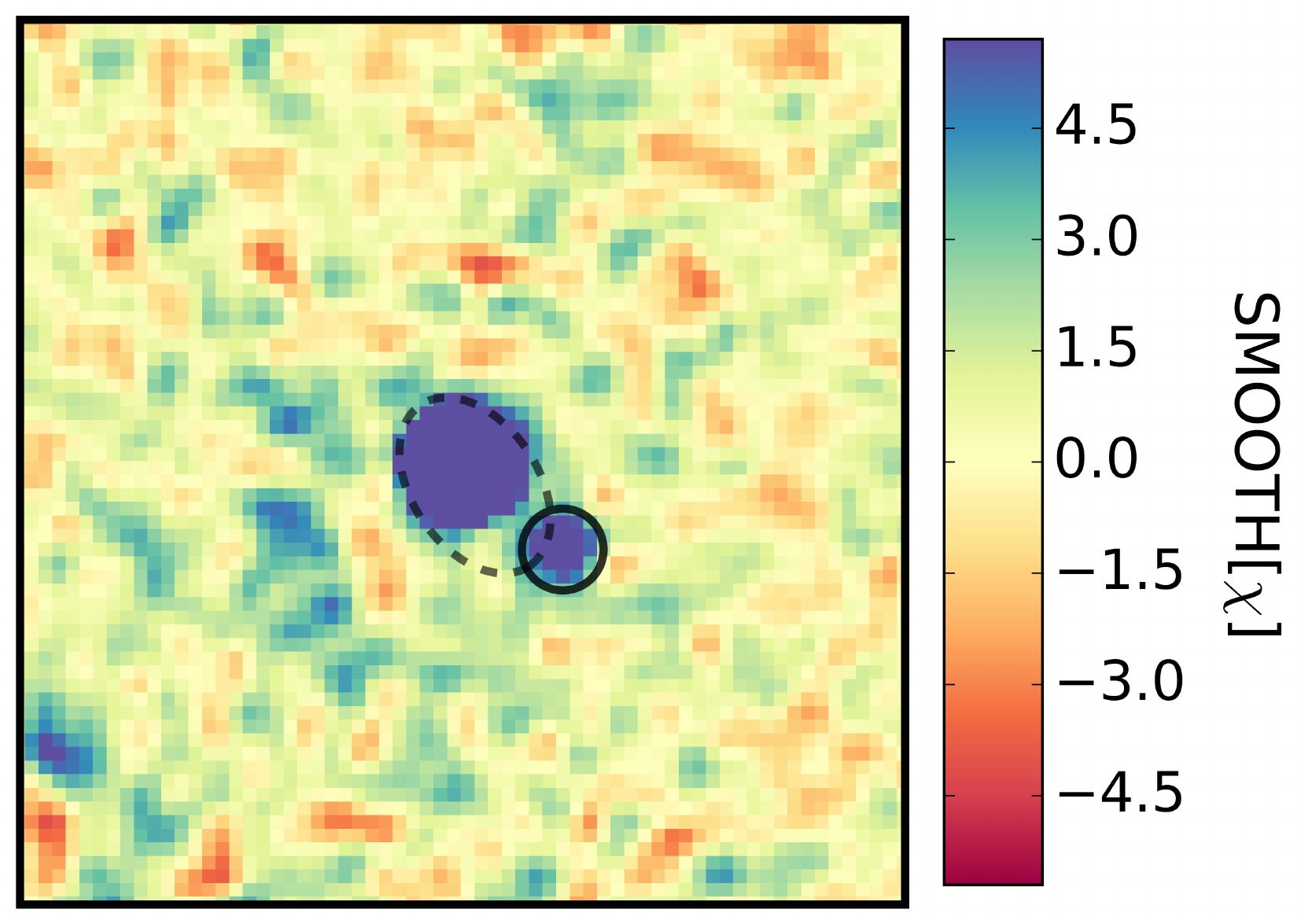}
\includegraphics[width=0.70\textwidth]{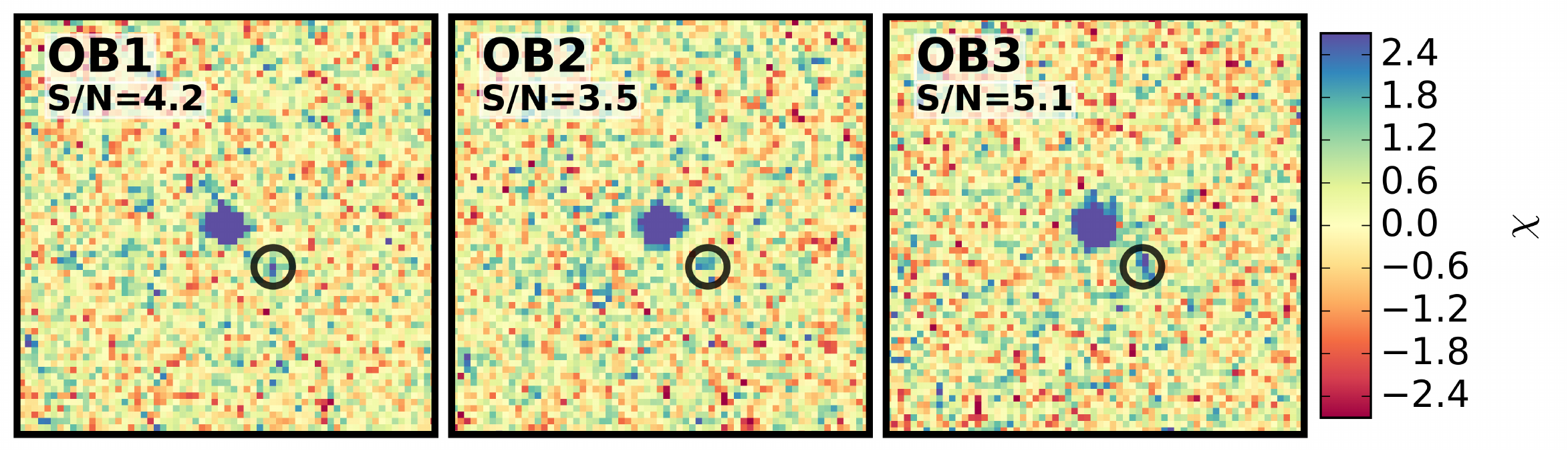}
\caption{
Detection of one LAE candidate in the proximity of \J0305. 
The different panels on the top show (from left to right) \mbox{${\rm DATA}_{i}$}, ${\mathlarger \chi}_{i}$, and ${\rm SMOOTH}\left[{\mathlarger \chi}_{i}\right]$ pseudo narrowband images obtained summing up the wavelength slices from 9269.75\,\AA\ to 9277.25\,\AA\ in the ${\rm DATA}_{\lambda,i}$ datacube.
A source (highlighted with black circles) is detected 2\farcs3 ($\sim$12.5\,kpc) from \J0305\ with a formal significance of 8.3--$\sigma$.
The dashed ellipse marks the position where the extended \lya\ halo is located (see Figure~\ref{fig:halo}).
On the bottom, pseudo narrowband ${\mathlarger \chi}_{i}$ images obtained from the single OBs exposures.
The source is detected with a SNR$>$3 in all the images.
Box sizes and color scales are the same as in Figure~\ref{fig:halo}.
}\label{fig:lae}
\end{center}
\end{figure*}

\begin{figure}[tb]
\begin{center}
\includegraphics[width=0.48\textwidth]{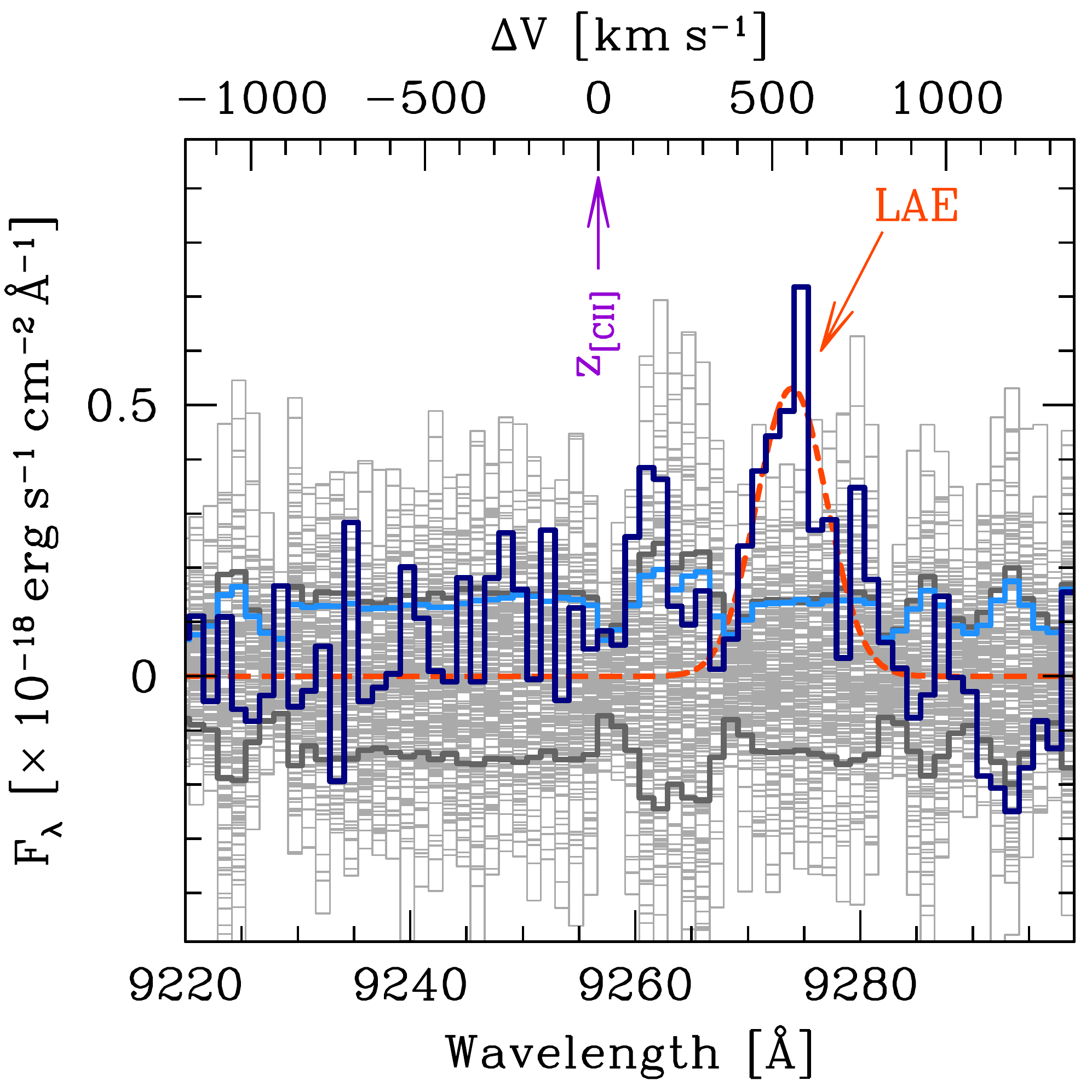}
\caption{
Spectrum of the LAE extracted over a 0\farcs6 circular aperture in the \mbox{${\rm DATA}_{\lambda,i}$} datacube.
The color code of the lines is the same as in Figure~\ref{fig:halospec}.
The detected line is well fitted by a Gaussian shifted 560\,km\,s$^{-1}$ with respect to the QSO's systemic redshift (estimated from the [\cii] emission line) and with a FWHM of~240\,km\,s$^{-1}$.
}\label{fig:laespec}
\end{center}
\end{figure}

\begin{figure}[tb]
\begin{center}
\includegraphics[width=0.48\textwidth]{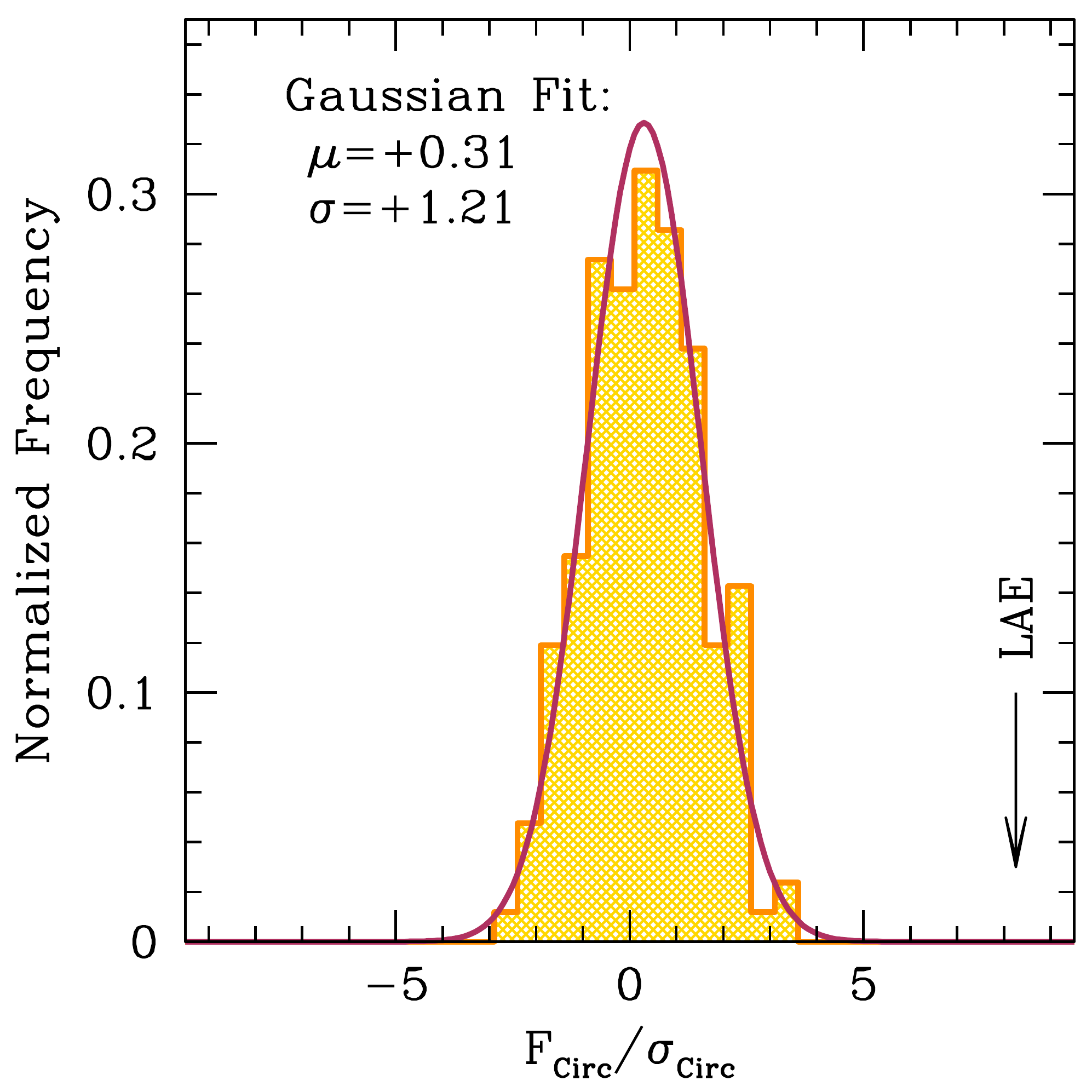}
\caption{
Analysis of the significance of the LAE detected in the MUSE datacube.
The histogram is created as for Figure~\ref{fig:halosyst} but with circular aperture with 0\farcs6 radius extracted over the collapsed \mbox{${\rm DATA}_{\lambda,i}$} datacube.
The Gaussian model of the distribution has $\mu$=0.31 and $\sigma$=1.21.
Correcting for systematic, the significance of the LAE is 6.8--$\sigma$.
}\label{fig:laesyst}
\end{center}
\end{figure}

\section{SEARCHING FOR LYMAN ALPHA EMITTERS}\label{sec:lae}

The rapid formation of SMBHs in the early Universe may imply that the first QSOs are tracers of galaxy overdensities.
In fact, the black hole growth may be fostered by rich environments, where interactions and mergers are more likely to occur \citep[see][ for a review]{Volonteri2012}.
To test this hypothesis we performed a search of LAEs in the proximity of \J0305. 

To identify LAEs associated with the QSO we focused our attention on the wavelength range between $\sim$9226\,\AA\ and$\sim$9287\,\AA, corresponding to $\pm$1000\,km\,s$^{-1}$ from the QSO systemic redshift.
This region was recursively sliced in 10\,\AA\ wide (i.e., $\sim$325\,km\,s$^{-1}$) pseudo narrowband images.
The nominal 5--$\sigma$ surface brightness limit reached in the image centered at 9256\,\AA\ is SB$_{5\sigma,\lambda}^1$=2.3$\times$10$^{-18}$\,erg\,s$^{-1}$\,cm$^{-2}$\,arcsec$^{-2}$ over a 1\,arcsec$^2$ aperture.
Assuming a spatially unresolved line emission with FWHM of 270\,km\,s$^{-1}$ \citep[i.e., the FWHM of the $z$$\sim$6.6 LAE spectral template presented in][]{Ouchi2010}, the corresponding 5--$\sigma$ luminosity limit is ${\rm L}_{5\sigma}$=6.7$\times$10$^{41}$\,erg\,s$^{-1}$.
Each image was then processed using \textsc{SExtractor} \citep{Bertin1996} requiring a minimum detection area of 7\,spaxel and a detection threshold of \mbox{1.5--$\sigma$}.
Identified sources were considered only if located within a 50\arcsec$\times$50\arcsec\ box centered on the QSO.
This search--box, smaller than the full MUSE field--of--view, was chosen to avoid issues related to the shorter exposure times experienced by the peripheral regions due to the dithering.
The neutral hydrogen in the intergalactic medium is expected to suppress virtually all the flux blueward of the \lya\ line in high redshift galaxies.
We thus excluded from our analysis sources present also in the pseudo narrowband image obtained collapsing  \mbox{${\rm DATA}_{\lambda,i}$} over the wavelength slices 4750\,\AA\,--\,8500\,\AA.
Finally, single exposure datacubes were visually inspected to identify false detections associated with highly deviating pixels present in only one OB.

This procedure allowed us to reveal the presence of a LAE in the immediate proximity of \J0305\ (RA=03:05:16.80, Dec.=-31:50:57.3, Epoch=J2000, see Figure~\ref{fig:lae}).
This appears as a 8.3--$\sigma$ detection in the pseudo narrowband images created summing up slices in the \mbox{${\rm DATA}_{\lambda,i}$} datacube from 9269.75\AA\ to 9277.25\AA, and as a 4.2, 3.5, 5.1--$\sigma$ detection in the single OB datacubes collapsed in the same wavelength range (see Figure~\ref{fig:lae}).
Figure~\ref{fig:laespec} shows the spectrum extracted over a circular aperture with a radius of 3\,spaxel.
By fitting a Gaussian function over the most prominent emission line, we derive a redshift of $z_{\rm LAE}$=6.629 (i.e. redshifted by $\sim$560\,km\,s$^{-1}$ with respect to the QSO's systemic redshift), a ${\rm FWHM}_{\rm LAE}$=240\,km\,s$^{-1}$, and a luminosity $\llya$=(2.1$\pm$0.2)$\times$10$^{42}$\,erg\,s$^{-1}$.
Consistently, the luminosity directly estimated from the pseudo narrowband image is $\llya$=(1.8$\pm$0.2)$\times$10$^{42}$\,erg\,s$^{-1}$.
No significant continuum emission is detected redward of the \lya\ line.

To quantify the effect of systematics we performed the same empirical test used in Section~\ref{sec:halo}.
From the pseudo narrowband image, we extracted an ensemble of circular apertures with 3\,spaxel radius, avoiding bright sources and image edges.
The distribution of the flux over noise ratio of these apertures is well reproduced by a Gaussian with average: $\mu$=0.31 and sigma: $\sigma$=1.21 (see Figure~\ref{fig:laesyst}).
The significance of the detection of the LAE emitter, once systematics are taken into account, is thus 6.8--$\sigma$.
In Figure~\ref{fig:laespec} we compare the spectra extracted from the circular apertures considered above with the spectrum of the LAE.
Despite the relatively low SNR per voxel of the line, 7~consecutive slices have a SNR$>$2 and the whole emission is the brightest line detected within $\pm$1000\,km\,s$^{-1}$ from the QSO's systemic redshift.
These results corroborate the reliability of the LAE's detection.
Given the high--redshift of this source, MUSE is not able to cover additional rest--frame UV line diagnostics (e.g., \heii, \civ) which would allow us to better determine the strength (and nature) of this faint companion.  

\section{DISCUSSION} \label{sec:discussion}

In the following we discuss the implications of the tentative detection of the extended \lya\ emission associated with the high redshift QSO \J0305\ (Sections~\ref{sec:host} and ~\ref{sec:cgm}) and of its companion galaxy (Section~\ref{sec:clust}).
It is worth mentioning that the real luminosity of the \lya\ emission could be slightly underestimated due to the presence of neutral hydrogen in the proximity of the QSO.
This effect is however negligible in the context of the forthcoming discussion.

\begin{figure}[tb]
\begin{center}
\includegraphics[width=0.48\textwidth]{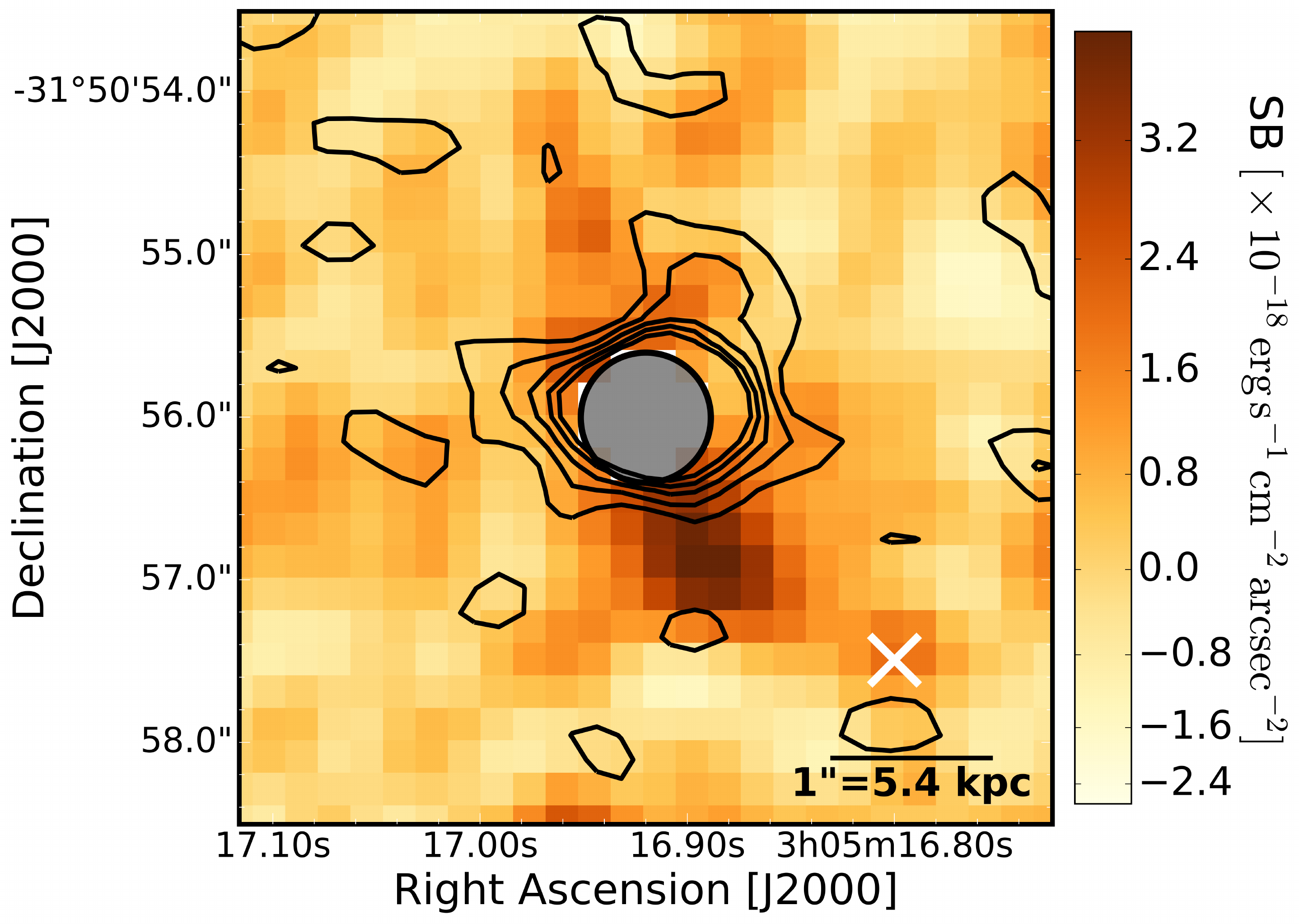}
\caption{
Comparison between the extended \lya\ emission detected in the PSF--subtracted MUSE datacube (background image, see also Figure~\ref{fig:halo}) and the (resolved) [\cii] emission line detected with ALMA by~\citet[][, black contours]{Venemans2016}.
The black contours trace the [\cii] line emission of \J0305\ at [1.5, 3.0, 4.5, 6.0, 7.5]--$\sigma$ significance level.
The PSF--subtracted pseudo narrowband image was convolved with a 2D Gaussian kernel with $\sigma$=1\,spaxel after removing the central region used to normalize the PSF (grey filled circle). 
The white cross on the bottom right corner marks the position of the LAE detected at $z$=6.629.
}\label{fig:alma}
\end{center}
\end{figure}

\begin{figure}[tb]
\begin{center}
\includegraphics[width=0.48\textwidth]{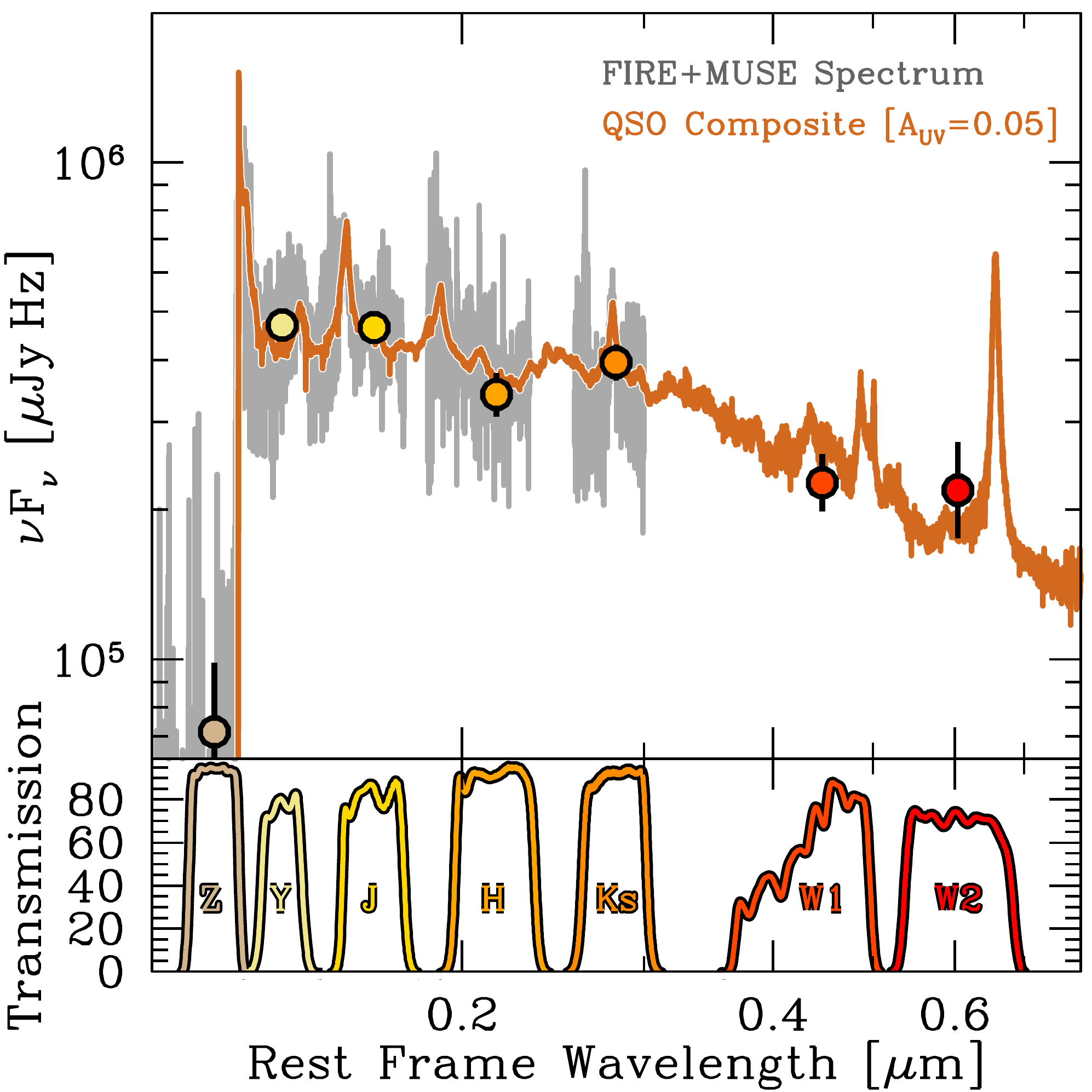}
\caption{
Top Panel --- 
SED of \J0305\ \citep[filled circles, see][ for details on the photometry]{Venemans2013}.
The combine FIRE$+$MUSE spectrum of the QSO is shown in grey \citep[see][ for details]{Derosa2014}.
The brown spectrum is the QSO composite from \citet[][]{Selsing2016} corrected for IGM absorption following \citet{Meiksin2006}.
The normalization is set by the W2 photometry.
A mild extinction of A$_{UV}$=0.05 permits to match the photometry in the bluer bands.
This imply that dust obscuration does not strongly affect the QSO emission along the line of sight.
Bottom Panel --- Transmission curves of the Z, Y, J, H, Ks, W1 and W2 bands used to create the SED.
}\label{fig:sed}
\end{center}
\end{figure}

\subsection{The Host Galaxy of \J0305}\label{sec:host}

In Section~\ref{sec:halo} we present a possible detection of extended \lya\ emission around the QSO \J0305.
The significance of this emission is strengthened by its spatial and redshift position (close to what is expected from the [\cii] emission).
In this section we link this discovery with the properties of the QSO's host galaxy.

Figure~\ref{fig:alma} maps the distribution of [\cii] (as observed with ALMA) overplotted to the \lya\ halo detected with MUSE.
The two emissions appear as not co--spatial.
A displacement of the \lya\ and [\cii]/dust emission has been commonly observed in other dusty sources at high redshift.
For instance, \citet{Hodge2015} reported a $\sim$4\,kpc offset between the rest--frame UV and both the FIR and the CO emission in the $z$$=$4.05 submillimeter galaxy GN20.
\citet{Decarli2016} and \citet{Aravena2016} observed a comparable shift in the compact star forming galaxy ASPECS~C.1 ($z$$=$2.54).
In a similar fashion, the \lya\ extended emission not associated with the radio jet, the rest--frame UV continuum, and the dust emission (albeit the coarse spatial resolution) appear to be offsetted in the $z$$=$4.11 radio galaxy TN~J1338$-$1942 \citep[e.g.][]{DeBreuck2004, Zirm2005, Venemans2007, Swinbank2015}.

For \J0305\ we can estimate the obscuration due to the copious amounts of dust detected with ALMA [${\rm M}_{\rm dust}$=(4.5--24)$\times$10$^8$\,${\rm M}_\odot$, \citealt{Venemans2016}] by comparing the luminosity of the observed extended \lya\ emission with the theoretical (unobscured) \lya\ emission expected due to the UV photons coming from the intense star--burst detected at millimeter wavelengths.
Assuming a case--B recombination, the relation between $\lha$ and SFR \citep[][]{Kennicutt2012} becomes:
\begin{equation}\label{eq:sfr}
\frac{\llya}{10^{42}{\rm erg}\,{\rm s}^{-1}}=1.62\frac{\rm SFR}{{\rm M}_\odot\,{\rm yr}^{-1}}.
\end{equation}
The star--formation rate of the host galaxy (SFR$_{\rm TIR}$=545\,$\msunyr$) yields to a \lya\ luminosity of $\llya$=8.8$\times$10$^{44}$\,erg\,s$^{-1}$, i.e. a factor $\sim$300$\times$ brighter than the observed luminosity.
This imply high extinction with $A_{\rm UV}$$\sim$6.2\,mag.

Resonance scattering may trap \lya\ photons and dim the expected luminosity of the extended emission.
The bouncing of \lya\ photons between optically thick clouds increases the total pathlength traveled in the dusty medium, incrementing its extinction (see Discussion in \citealt{Decarli2012} and Appendix in \citealt{Hennawi2013}).
The relative importance of this effect strongly depends on various QSO's host galaxy properties, such as, among the others, neutral hydrogen column density, neutral fraction, dust--to--gas ratio, and geometry.
However, there is no clear evidence for a broad double peaked kinematics as expected from resonantly trapped \lya\ photons \citep[e.g.,][]{Cantalupo2005}, possibly due to absorption from neutral hydrogen in the intergalactic medium.
In addition, an high level of obscuration is in contrast with the low extinction observed toward the QSO:
Figure~\ref{fig:sed} plots broad--band photometry for \J0305\ taken from \citet{Venemans2013}. 
The SED matches the QSO composite spectrum of \citet[][]{Selsing2016} well, leaving room for only little extinction.
Assuming a SMC extinction curve \citep{Gordon2003}, appropriate for non--BAL QSOs at high redshift \citep[e.g.,][]{Gallerani2010}, we can infer a stringent limit on the obscuration of $A_{\rm UV}$$<$0.1\,mag.

These results suggest that the copious amounts of dust detected with ALMA [${\rm M}_{\rm dust}$=(4.5--24)$\times$10$^8$\,${\rm M}_\odot$, \citealt{Venemans2016}] can effectively obscure the UV emission and can prevent the ionizing photons from escaping the QSO's host galaxy.
Yet, UV radiation is able to leak through small openings in the dust cocoon.
These leakage may be associated with short and extreme bursts of star formation as proposed to explain the intrinsically blue rest frame UV slope observed in dusty, star forming galaxies (SFR$>$50\,M$_\odot$\,yr$^{-1}$) up to $z$$\sim$5 \citep[e.g.,][]{Casey2014}. 
A similar scenario was also proposed by \citet{Decarli2012} to explain the lack of extended \lya\ emission from the host galaxy of the two highly star forming $z$$>$6 QSOs: SDSS~J1030+0524 and SDSS~J1148+5251 \citep[see also][]{Mechtley2012}.

\begin{figure}[tb]
\begin{center}
\includegraphics[width=0.48\textwidth]{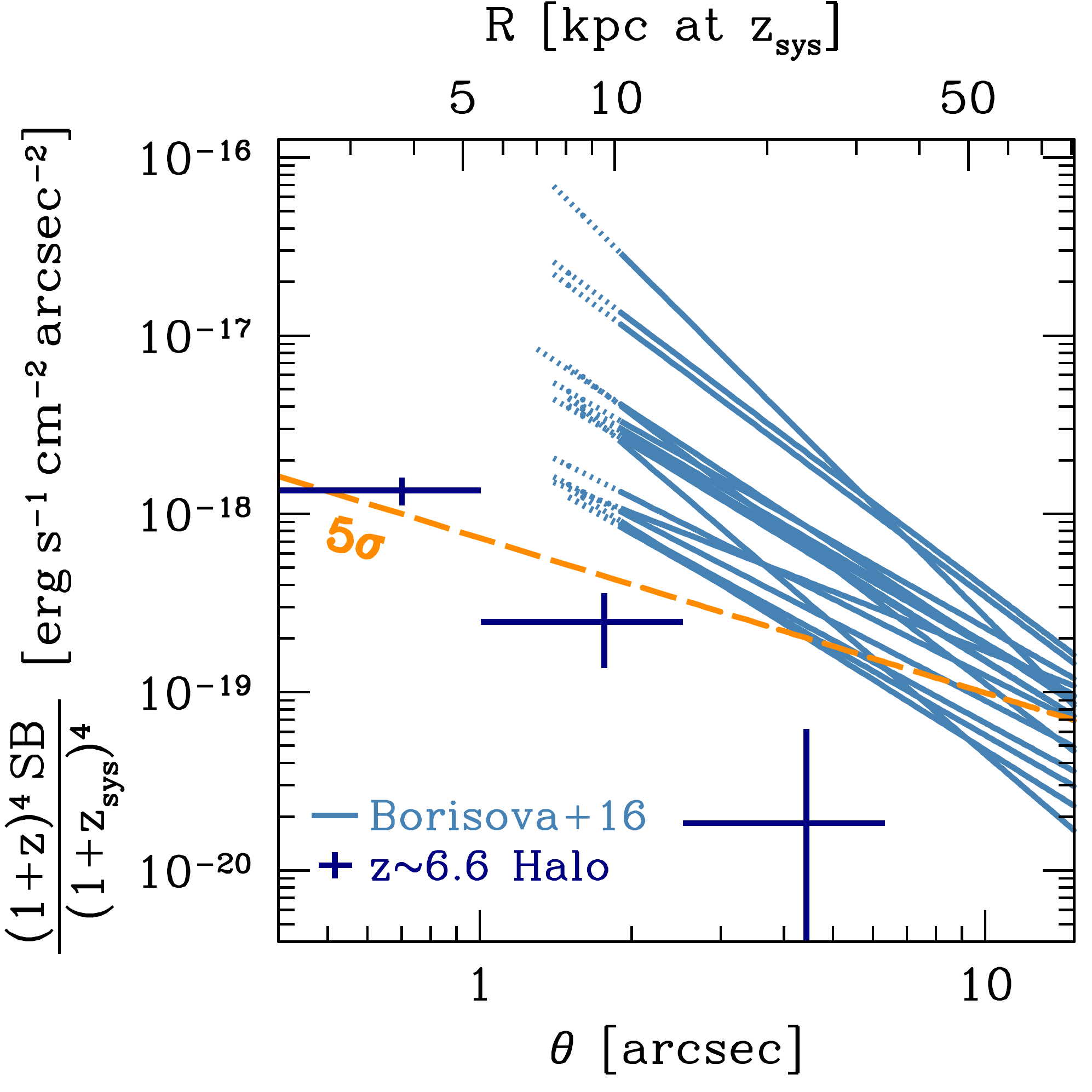}
\caption{
Redshift--corrected surface brightness radial fit of $z$$\sim$3.5 radio quiet QSOs from \citet[][, light blue solid lines]{Borisova2016} compared with the nominal 5--$\sigma$ surface brightness limit reached by collapsing the datacube over 5 wavelength slices centered at $z_{\rm sys}$ (orange dashed line).
The power--law fits for intermediate redshift QSOs are extrapolated down to a separation of 1\arcsec\ (at the QSO's redshifts) where no information on the extended emission could be provided due to the PSF subtraction procedure \citep[light blue dotted lines, see][ for details]{Borisova2016}.
The circular averaged profile of emission around \J0305\ is also showed for comparison (dark blue crosses).
This profile is calculated in annuli with radii evenly spaced in logarithmic space.
Despite its asymmetric morphology, the tentative extended emission is detected just above the 5--$\sigma$ level in the inner 0\farcs4--1\farcs0 annulus.
}\label{fig:borisova}
\end{center}
\end{figure}

\begin{figure}[tb]
\begin{center}
\includegraphics[width=0.48\textwidth]{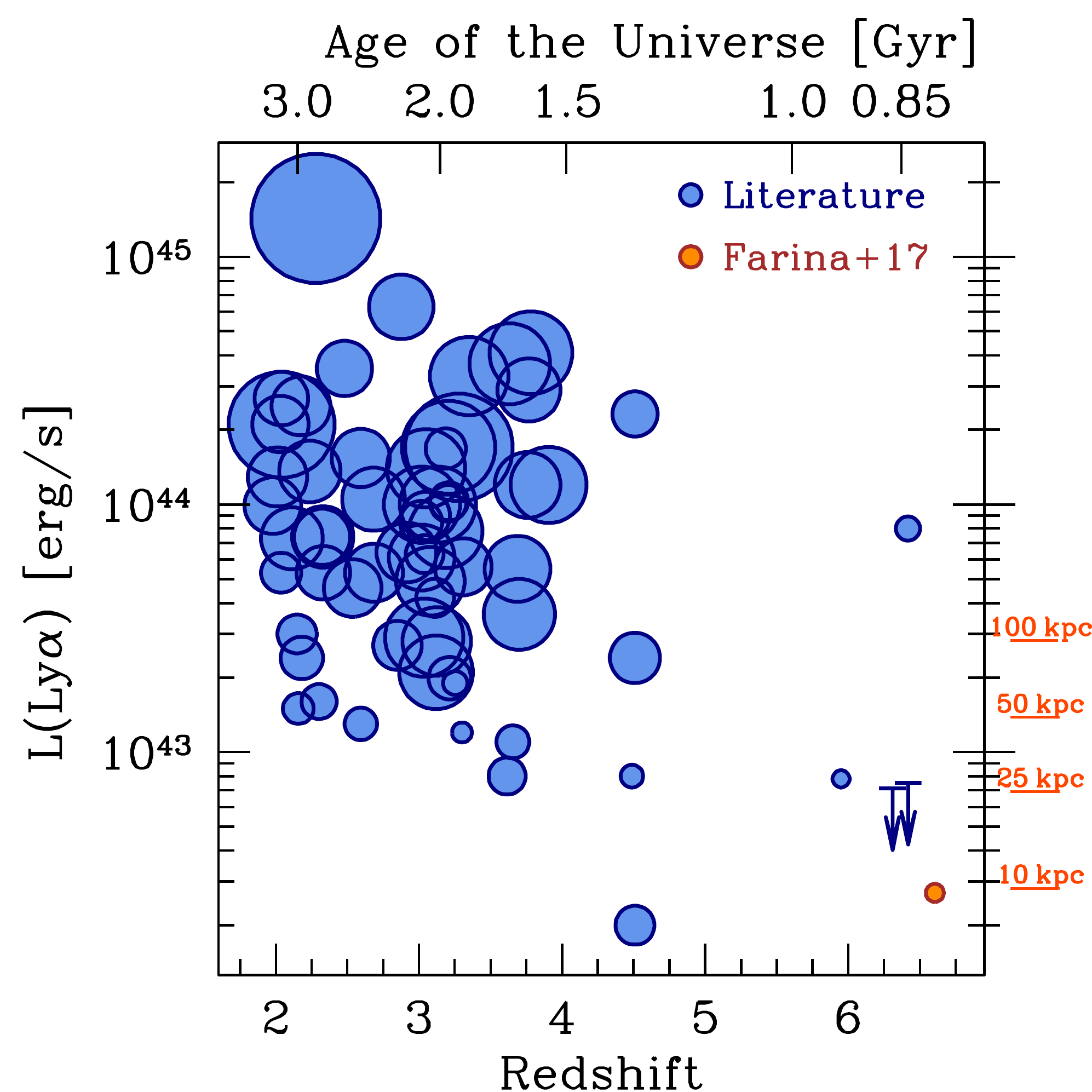}
\caption{
Distribution of all \lya\ nebulae associated with QSOs know to date in the redshift vs. total \lya\ luminosity plane (blue circles). 
The radius of the points is proportional to square root of the maximum extent of the \lya\ emissions.
Data at $z$$>$5 are SDSS~J2228+0110 \citep[$z$=5.95,][]{Roche2014} and CFHQS~J2329$-$0301 \citep[$z$=6.42,][]{Willott2011}
The arrows are 5--$\sigma$ upper limits set by \citet{Decarli2012} on SDSS~J1030+0524 ($z$=6.31) and SDSS~J1148+5251 ($z$=6.42, see text for details).
The extended emission associated with \J0305\ is plotted as an orange filled circle.
Ticks on the right side of the plot mark nominal sensitivity limits reached in the MUSE data calculated assuming SB$_{5\sigma}^1$=1.9$\times$10$^{-18}$\,erg\,s$^{-1}$\,cm$^{-2}$\,arcsec$^{-2}$ over a 1\,arcsec$^2$ aperture (obtained collapsing the cube over 5 wavelength slices, see Section~\ref{sec:lim}) and top hat sources with diameters 10, 25, 50, and 100\,kpc and a FWHM of 200\,km\,s$^{-1}$.
}\label{fig:red_lum}
\end{center}
\end{figure}

\subsection{Large Scale Ly$\alpha$ Emission}\label{sec:cgm}

Is the surface--brightness limit reached with MUSE sufficient to probe the presence of \lya\ nebular emission on scales of 5--100\,kpc?
To test this possibility we compare our detection limits with the circular averaged surface brightness profiles of bright QSOs at intermediate redshift.
As comparison sample we consider the recent work of \citet{Borisova2016}, who investigated 17~radio quiet QSOs at 3.0$\lesssim$$z$$\lesssim$3.9.
We stress that, with a median absolute magnitude at 1450\,\AA\ of $\M1450$$=$$-29.2$, these are among the brightest QSOs at $z$$\sim$3.5, hence they are $\sim$3 magnitudes brighter than \J0305.
The nebulae surrounding these QSOs extend on scales of 50--100\,kpc and their circularly average surface brightness profiles are well represented by a power--law decline in the majority of the cases.
In Figure~\ref{fig:borisova} we plot the power--law best fits of these extended \lya\ emissions as they would appear if moved at $z$=$z_{\rm sys}$, which means we corrected for redshift dimming and for different angular diameter distances.
We also show the profile of the extended \lya\ emission associated with \J0305\ averaged over annuli evenly spaced in logarithmic space together with our nominal 5--$\sigma$ limit on the surface brightness.
It is apparent that if \J0305\ was surrounded by an extended emission similar to bright $z$$\sim$3.5 QSOs, our MUSE data would have been deep enough to detect it at angular scales $\theta$$\gtrsim$1\arcsec.

The nebular emission around \J0305\ appears to have a significantly lower surface brightness and lower total luminosity than commonly observed at intermediate redshift. 
In Figure~\ref{fig:red_lum} we present a compilation of \lya\ nebulae detected around QSOs from the literature\footnote{Data are from: \citet{Heckman1991a, Heckman1991b, Bremer1992, Roettgering1997, vanOjik1997, Lehnert1998, Bergeron1999, Fynbo2000, Bunker2003, Weidinger2004, Weidinger2005, Christensen2006, Courbin2008, Barrio2008, Yang2009, Smith2009, Matsuda2011, Willott2011, North2012, Humphrey2013, Cantalupo2014, Roche2014, Husband2015, Hennawi2015, Borisova2016, Fumagalli2016, Fathivavsari2016}. Effects of different sensitivities are not taken into account.
}.
5--$\sigma$ upper limits on the luminosity set by \citet{Decarli2012} on SDSS~J1030+0524 ($z$=6.31) and SDSS~J1148+5251 ($z$=6.42) are also plotted.
These are derived rescaling their 5--$\sigma$ surface brightness limit (SB$_{5\sigma}^1$=1.0$\times$10$^{-17}$\,erg\,s$^{-1}$\,cm$^{-2}$\,arcsec$^{-2}$ over a 1\,arcsec$^2$ aperture) to a circular aperture with a diameter of 10\,kpc.
On the right side of Figure~\ref{fig:red_lum} we mark the 5--$\sigma$ upper limits on the luminosity for top hat sources with diameters d=[10, 25, 50, 100]\,kpc and FWHM=200\,km\,s$^{-1}$.
These are calculated considering the surface brightness limit obtained collapsing the MUSE datacube over 5~wavelength slices (SB$_{5\sigma,\lambda}^1$=1.9$\times$10$^{-18}$\,erg\,s$^{-1}$\,cm$^{-2}$\,arcsec$^{-2}$ over a 1\,arcsec$^2$ aperture, see Section~\ref{sec:lim}).
These data are suggestive for a decline of total luminosity of the \lya\ nebulae as a function of redshift.
This may indicate of a change in the gas properties and/or in the powering mechanisms at different epochs. 
A more quantitative interpretation, however, is hampered by differing methodologies, ambiguities in detection criteria, and the lack of a statistical sample of QSOs investigated at $z$$>$4.

It is of interest to compare our observational results with predictions on the luminosity of the \lya\ for different emission mechanisms. 
In Section~\ref{sec:thick} we consider the case of recombination from optically thick clouds.
The case in which the QSO radiation highly ionize the surrounding gas that thus become optically thin is addressed in Section~\ref{sec:thin}.
Finally, in Section~\ref{sec:mix} we comment on the possibility that the gas is in a multi--phase status. 
These calculations closely follow the formalism described in \citet{Hennawi2013}.

\subsubsection{Optically Thick Gas}\label{sec:thick}

Under the assumption that the surrounding of the QSO is filled by cool, optically--thick clouds, self--shielding generates a thin, highly--ionized envelope around individual clouds that acts as a mirror converting a fraction of the ionizing radiation into \lya\ photons \citep[][]{Gould1996}.
In this scenario the powering mechanism is the QSO ionizing radiation (${\rm L}_{\nu_{\rm LL}}$, where  $\nu_{\rm LL}$ is the frequency at the Lyman edge).
$\llya$ is thus proportional to ${\rm L}_{\nu_{\rm LL}}$ as:
\begin{equation}\label{eq:fluo}
\frac{\llya}{10^{44}\,{\rm erg}\,{\rm s}^{-1}}=
7.8\,f^{\rm thick}_{c}\frac{{\rm L}_{\nu_{\rm LL}}}{10^{30}\,{\rm erg}\,{\rm s}^{-1}\,{\rm Hz}^{-1}}
\end{equation}
where $f^{\rm thick}_{c}$ is the optically thick clouds covering factor.
If we assume $f^{\rm thick}_{c}$=0.1 as estimated for the small scale \lya\ emission observed in $z$=2--3 QSOs by \citet[][]{Hennawi2013} and ${\rm L}_{\nu_{\rm LL}}$=$4.9$$\times$10$^{30}$erg\,s$^{-1}$\,Hz$^{-1}$ (obtained rescaling the composite spectrum from \citet{Lusso2015} to the QSO luminosity at 1350\,\AA) we obtain $\llya$=3.8$\times$10$^{44}$\,erg\,s$^{-1}$.
This discrepancy of a factor $\sim$130$\times$ with respect to the observed luminosity may be due to the geometry of the emission:
If the UV photons break only through a small solid angle $\Omega_{\rm e}$, the expected \lya\ luminosity would be reduced by a factor $f_{\Omega_{\rm e}}$$=$$\left(\frac{\Omega_{\rm e}}{4\pi}\right)$.
Assuming, for the sake of simplicity, that the anisotropic emission occurs in a cone, an opening angle of 30\textdegree\ corresponds to $f_{\Omega_{\rm e}}$$\sim$0.07.
However, this would imply an unrealistic fraction of obscured AGN of $f_{\rm Obs}$$=$$\left(1-\frac{\Omega_{\rm e}}{4\pi}\right)=0.85$ \citep[e.g.][]{Treister2008, Lusso2013, Merloni2014}.
Alternatively, a factor 100$\times$~lower in the optically thick clouds covering fraction may explain the faintness of the emission. 
Such a low covering fraction however is in contrast with results from $z$$\sim$2--3 QSOs \citep[][, see also Section~\ref{sec:mix}]{Prochaska2013a}.
This result push for different emission mechanisms to explain the observed emission.

\subsubsection{Optically Thin Gas}\label{sec:thin}

If the gas surrounding the QSO is optically thin, the QSO radiation would be sufficiently intense to keep the gas highly ionized (i.e., the hydrogen neutral fraction is $x_{\rm HI}$=$\frac{n_{\rm HI}}{n_{\rm H}}$$\ll$1).
As shown in \citet[][]{Hennawi2013}, in the optically thin regime $\llya$ can be expressed in terms of the area--averaged neutral column density ($\langle N_{\rm H\,I} \rangle$) and of the ionizing luminosity:
\begin{equation}\label{eq:thin}
\frac{\llya}{10^{44}\,{\rm erg}\,{\rm s}^{-1}}=
0.9\,\frac{\langle N_{\rm H\,I} \rangle}{10^{17.2}\,{\rm cm}^{-2}}
\frac{{\rm L}_{\nu_{\rm LL}}}{10^{30}\,{\rm erg}\,{\rm s}^{-1}\,{\rm Hz}^{-1}}
\end{equation}
where the normalization of $\langle N_{\rm H\,I} \rangle$ is set by the requirement that, to be optically thin, clouds must have $N_{\rm H\,I}$$\ll$10$^{17.2}$\,cm$^{-2}$. 
Plugging in this equation the observed luminosity of the nebula and the ${\rm L}_{\nu_{\rm LL}}$ estimated above we obtain $\langle N_{\rm H\,I} \rangle$$\sim$10$^{15.0}$\,cm$^{-2}$.
It is thus plausible that the extended emission arises from optically--thin clouds illuminated by the QSO.
However, it is worth to remind that while $\langle N_{\rm H\,I} \rangle$$>$10$^{17.2}$\,cm$^{-2}$ implies that the gas is in the optically thick regime, clouds with $\langle N_{\rm H\,I} \rangle$$<$10$^{17.2}$\,cm$^{-2}$ could be either optically thin or optically thick \citep[see][]{Hennawi2013}.

If we assume an optically thin scenario, we can relate the observed \lya\ surface brightness with the hydrogen total column density ($N_{\rm H}$) and volume density \citep[$n_{\rm H}$; e.g.,][]{Hennawi2013}:
\begin{equation}\label{eq:sbthin}
\begin{split}
\frac{\rm{SB}({\rm Ly}\alpha)}{10^{-19}\,{\rm erg}\,{\rm s}^{-1}\,{\rm cm}^{-2}\,{\rm arcsec}^{-2}}&=5.9\,\left(\frac{1+z}{7.6145}\right)^{-4}\\
\times f^{\rm thin}_c\,\left(\frac{n_{\rm H}}{1.0\,{\rm cm}^{-3}}\right)&\,\left(\frac{N_{\rm H}}{10^{20.5}\,{\rm cm}^{-2}}\right)
\end{split}
\end{equation}
where $f^{\rm thin}_{c}$ is the covering fraction of the optically thin gas.
From this equation, we can derive an estimate of the volume density of the gas giving rise to the extended emission.
In fact, the total column density of the hydrogen in proximity of $z$$\sim$2 QSOs has been constrained from the study of absorption systems.
Photoionization models of these absorbers suggest that $N_{\rm H}$ is almost constant within an impact parameter of 200\,kpc with a median value $N_{\rm H}$=10$^{20.5}$\,cm$^{-2}$ \citep[e.g.][]{Prochaska2009, Hennawi2015, Lau2016}.
Assuming $f^{\rm thin}_{c}$=0.5, to explain the observed surface brightness $\rm{SB}({\rm Ly}\alpha)$=(1.8$\pm$0.2)$\times$10$^{-18}$\,erg\,s$^{-1}$\,cm$^{-2}$\,arcsec$^{-2}$ (calculated over the elliptical aperture considered in Section~\ref{sec:halo}) an high gas volume density of $n_{\rm H}$=6.1\,cm$^{-3}$ is required.
Remarkably, smiliarly high $n_{\rm H}$ were proposed to explain the emission of the giant \lya\ nebulae associated with the QSOs UM~287 \citep{Cantalupo2014, Arrigoni2015giant} and SDSS~J0841+3921 \citep{Hennawi2015}.

\subsubsection{A Multi--Phase Scenario}\label{sec:mix}

In the previous sections, we estimated the expected emission from the gas surrounding \J0305\ considering that it is either optically thick or optically thin to the QSO radiation.
However, we can also consider a multi--phase scenario where low density clouds with high covering fraction are responsible for the absorption systems observed at $z$$\sim$2, whereas the observed emission rises from optically thick gas with low covering fraction (and hence rarely intercepted in absorption).
We also stress that disentangling among different emission mechanism is challenging.
At a given separation from the QSO, optically thick and optically thin gas clouds could result in a similar emission with opportune combinations of $n_{\rm H}$, $N_{\rm H}$, and covering fraction.
From Equations~\ref{eq:thin}~and~\ref{eq:sbthin} it is clear that, for a given $N_{\rm H}$, one can consider to increase $n_{\rm H}$ to rise the emission in the optically thin regime.
However, when the area--averaged neutral column density reach $\langle N_{\rm H\,I} \rangle$$\sim$10$^{17.2}$\,cm$^{-2}$, the cloud become optically thick and recombination would occur only in the self shielding layer.
We can thus roughly calculate the density required for a cloud to become optically thick as a function of the distance from the QSO by matching the emission estimates in Section~\ref{sec:thick} and \ref{sec:thin} (for the same covering factor).
We obtain that, in order to shelf shield at a separation of 25\,kpc from \J0305, a cloud should have a high density of $n_{\rm H}$$\sim$200\,cm$^{-3}$.

\subsection{Overdensity of LAE around the \J0305}\label{sec:clust}

The first search for LAEs around $z$$>$5.5 QSOs was performed by \citet{Decarli2012} using a combination of narrowband filters of the WFC3 on {\it HST}.
No companion galaxies were found around the two QSOs SDSS~J1030+0524 and SDSS~J1148+5251.
However, the field--of--view of WFC3 allowed the authors to probe only a small cosmological volume and a relatively bright point source detection limit was reached ($\sim$23.4\,mag).
\citet{Banados2013} and \citet{Mazzucchelli2016} used sensitive narrow-- and broad--band images from the FOcal Reducer/low dispersion Spectrograph~2 (FORS2) on the VLT to investigate the environment of the two QSOs ULAS J0203+0012 ($z$=5.72) and PSO~J215.1512-16.0417 ($z$=5.73) over a much larger area ($\sim$37\,arcmin$^2$).
Both studies report a number of LAEs consistent with (or even lower than) expectations from a blank field.
\citet{Goto2017}, using the Subaru Prime Focus Camera (Suprime--Cam) on the Subaru telescope, similarly reported an under--density of LAEs around the QSO CFHQS~J2329$-$0301 ($z$=6.4).
It is worth mentioning that at slightly lower redshift \citet{McGreer2014} discovered a LAE in the immediate proximity (with a separation of only 11.4\,kpc) of the $z$$\sim$4.9 QSO SDSS~J0256+0019.

The detection of one LAE in the MUSE datacube at only $\sim$12.5\,kpc and 560\,km\,s$^{-1}$ from \J0305\ may represent the first spectroscopically confirmed evidence of the connection between high--density environments (and thus high merger rate) and $z$$>$6 QSOs, that was postulated to explain the rapid assembly of the first SMBHs \citep[e.g.][]{Volonteri2012}.
To confirm this scenario we have first to calculate the probability of finding such LAE in the proximity of the QSO.

\begin{figure}[ht]
\begin{center}
\includegraphics[width=0.48\textwidth]{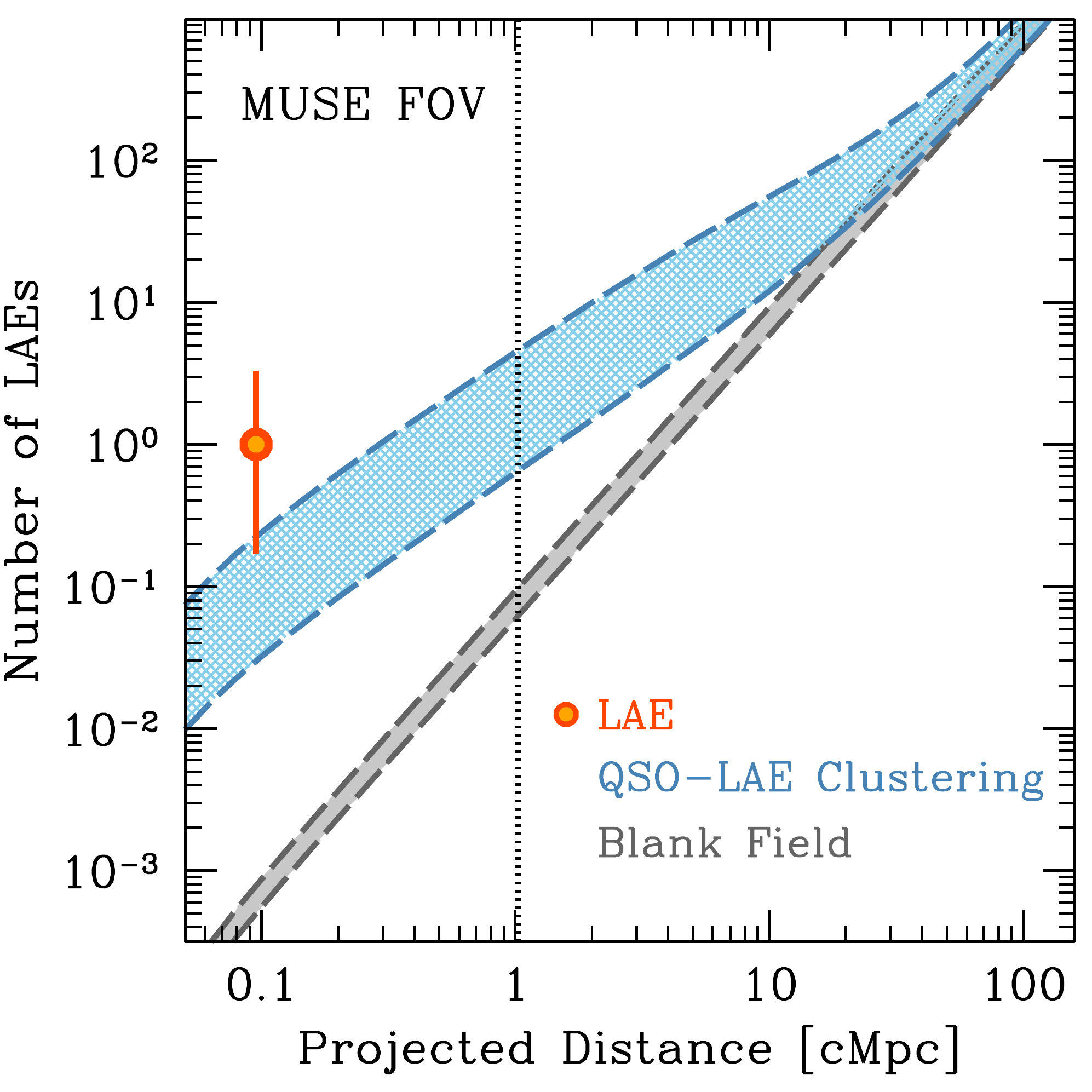}
\caption{
Number density of LAEs as a function of Projected Distance.
The grey shaded area shows the expected number of LAEs located within $\pm$1000\,km\,s$^{-1}$ from the QSO systemic redshift.
This is obtained integrating the $z$$\sim$6.6 LAE luminosity function of \citet{Matthee2015} down to the 5--$\sigma$ luminosity limit reached by our observations (i.e., ${\rm L}_{5\sigma}$=10$^{41.8}$\,erg\,s$^{-1}$).
In absence of clustering, the LAE detected at a projected distance of 12.5\,kpc from \J0305\ (orange point, error bars are the 1--$\sigma$ confidence interval derived following \citealt{Gehrels1986}) appears to be a factor $\sim$1000$\times$ above expectations.
The large scale QSO--LAE clustering (see Section~\ref{sec:clust}) increase the expected number of LAE in the proximity of a QSO (blue shaded area).
Even considering the upper--limit on the $z$$\sim$6.6 LAE--LAE auto--correlation function to determine the QSO--LAE cross--correlation length less then $\sim$0.1 LAEs are expected at separation $<$15\,kpc.
}\label{fig:clust}
\end{center}
\end{figure}

\subsubsection{Comparison to Blank Field}

In order to estimate how many LAEs are expected from the blank field we integrated the  $z$$\sim$6.6 LAE luminosity function from \citet{Matthee2015}\footnote{As suggested in \citet{Matthee2015}, we conservatively consider the fit of the spectroscopically confirmed UDS+COSMOS sources with a fixed faint--end slope $\alpha$=$-1.5$.} down to our 5--$\sigma$ luminosity limit for an unresolved source (i.e., ${\rm L}_{5\sigma}$=6.7$\times$10$^{42}$\,erg\,s$^{-1}$, see Section~\ref{sec:lae}).
The derived number density of LAEs is ${\mathlarger \phi}$$($L$>$10$^{41.8}$erg\,s$^{-1}$$)$=(5.8$^{+1.9}_{-1.2}$)$\times$10$^{-4}$\,cMpc$^{-3}$.
This means that $\lesssim$0.1 LAEs are expected within the total volume explored in our analysis (i.e., 50\arcsec$\times$50\arcsec$\times$2000\,km\,s$^{-1}$, or $\sim$80\,cMpc$^3$, see Section~\ref{sec:lae}) and a mere $\sim$10$^{-3}$ within a separation of 12.5\,kpc (see Figure~\ref{fig:clust}).
Such a low incidence of LAEs due to the blank field supports the idea that the detected LAE has to be physically linked to the presence of the QSO.

\subsubsection{Comparison to QSO--LAE clustering}

The study of the QSO--galaxy and QSO--QSO clustering showed that bright QSOs at low and intermediate redshifts are biased tracer of massive dark matter haloes with M$_{\rm DM}$$\gtrsim$10$^{12.5}$\,M$_\odot$ \citep[e.g.,][]{Myers2007a, Shen2007, Padmanabhan2009, Ross2009, Farina2011, White2012, Richardson2012, Shen2013, Zhang2013, Sandrinelli2014, Karhunen2014, Eftekharzadeh2015, Eftekharzadeh2017, Garcia2017}.
At high redshift, constraints are given by the discovery of a QSO pair with projected separation of only~130\,kpc at $z$=5.02 \citep[][]{McGreer2016}.
The inferred correlation length is $r_0$$>$29\,cMpc, that is consistent with the $r_0$$\sim$30\,cMpc estimated by \citet[][]{Shen2007, Shen2010} from a sample of QSO pairs at 3.5$<$$z$$<$4.5.
To predict the expected number of LAEs in presence of clustering, we follow the formalism proposed by \citet{Hennawi2006b}.
In summary, the real space two point QSO--LAE cross--correlation function $\xi^{\rm QG}$ is integrated along the line--of--sight to eliminate distance distortions in the redshift space, and the expected number of companions is calculated in cylindrical shells (V$_{\rm shell}$) centered on the QSO.
In practice, $\xi_{\rm QG}$ can be expressed as:
\begin{equation}
\xi^{\rm QG}(r,R)=\left[\frac{\left(r^2+R^2\right)^{\frac{1}{2}}}{r_0^{\rm QG}}\right]^{-\gamma}
\end{equation}
where $r$ and $R$ are comoving coordinates, perpendicular to and along the line--of--sight, respectively; $r_0^{\rm QG}$ is the QSO--LAE cross--correlation length; and $\gamma$ is the slope of the correlation function \citep[e.g.][]{Peebles1980}.
In presence of clustering, the number of LAEs expected within a cylindrical shell of volume
${\rm V}_{\rm shell}$($r$)=$\pi$($r^{2}$$-$$r_{\rm min}^2$)$\Delta R$, where $r_{\rm min}$ is the minimum radius (in comoving Mpc) that can be probed in the MUSE datacube (i.e., the seeing radius), $r$ is the comoving transverse distance from the QSO, and $\Delta R$ is the comoving line--of--sight distance corresponding to the velocity range $\pm$1000\,km\,s$^{-1}$ from the QSO's systemic redshift, is:
\begin{equation}
{\rm N}_{\rm C}({\rm L}_{\rm Lim},r)={\mathlarger \phi}({\rm L}_{\rm Lim}){\rm V}_{\rm shell}(r)\left[1+\bar{{\rm W}}_{p}^{\rm QG}(r_{\rm min},r)\right]
\end{equation}
where the volume--averaged projected cross--correlation function ($\bar{{\rm W}}_{p}^{\rm QG}$) is given by:
\begin{equation}
\bar{{\rm W}}_{p}^{\rm QG}(r_{\rm min},r)=
\frac{
\int_{-\frac{\Delta R}{2}}^{+\frac{\Delta R}{2}}dR^\prime
\int^{r}_{r_{\rm min}} 2\pi r^\prime dr^\prime \xi^{\rm QG}(r^\prime,R^\prime)
}{{\rm V}_{\rm shell}(r)}.
\end{equation}
To infer $r_0^{\rm QG}$ we consider that, by definition, $\xi_{\rm QG}$ could be expressed as:
\begin{equation}\label{eq:xi}
\xi^{\rm QG}(r,R)=\left\langle\delta_{\rm Q}(r,R)\delta_{\rm G}(r,R)\right\rangle
\end{equation}
where $\delta_{\rm Q}$ and $\delta_{\rm G}$ are the relative density contrasts of QSOs and LAEs, respectively; and the angular brackets denote averaging over a distribution. 
Under the assumption that QSOs and LAEs trace the same underlying dark matter distribution and considering a linear bias ($\delta_{\rm Q}$=$b_{\rm Q}$$\delta_{\rm DM}$ and $\delta_{\rm G}$=$b_{\rm G}$$\delta_{\rm DM}$), Equation~\ref{eq:xi} can be rewritten as:
\begin{equation}
\begin{split}
\xi^{\rm QG}(s,R)&=b_{\rm Q}b_{\rm G}\xi^{\rm DMDM}(r,R)\\
                 &=\left[\xi^{\rm QQ}(r,R)\xi^{\rm GG}(r,R)\right]^{\frac{1}{2}}
\end{split}
\end{equation}
where $\xi^{\rm DMDM}$, $\xi^{\rm QQ}$, and $\xi^{\rm GG}$ are the real space two point auto--correlation functions for dark matter halos, QSOs, and LAEs, respectively \citep[e.g.][]{Schneider2006}.
Assuming a power law form for both $\xi^{\rm QQ}$ and $\xi^{\rm GG}$ (with the same~$\gamma$) the QSO--LAE cross--correlation length could be estimated as:
\begin{equation}
r_{0}^{\rm QG}=\left(r_{0}^{\rm QQ}r_{0}^{\rm GG}\right)^{\frac{1}{2}}
\end{equation}
where $r_{0}^{\rm QQ}$ and $r_{0}^{\rm GG}$ are the QSO--QSO and LAE--LAE auto--correlation lengths.
As a proxy for $r_{0}^{\rm QG}$ at $z$$\sim$6.6 we considered $r_{0}^{\rm GG}$=10.3$^{+4.7}_{-8.6}$\,cMpc that is the upper limit on the auto--correlation length derived by \citet{Ouchi2010} from the clustering analysis of bright LAEs at $z$=6.6 and $r_{0}^{\rm QQ}$=17.4$^{+2.5}_{-2.8}$\,cMpc obtained imposing $\gamma$=1.8 to the study of the clustering properties of $z$$>$2.9 QSOs of \citet{Shen2007}.
It is worth noticing that, at $z$$\sim$1--2, the QSO--QSO auto--correlation function appears to get steeper at sub--Mpc separations \citep[e.g.][]{Djorgovski1991, Hennawi2006b, Myers2007b, Djorgovski2007, Myers2008, Hennawi2010, Kayo2012, Farina2013trip}.
However, this enhancement become less prominent at $z$$>$2.9 where small-- and large--scale clustering amplitude are comparable \citep[]{Shen2010}.

Figure~\ref{fig:clust} shows, as a function of $r$, the estimated number of LAEs located within $\pm$1000\,km\,s$^{-1}$ from the QSO's systemic redshift.
Considering the cross--correlation length estimated above $\lesssim$0.1 LAEs are expected within 100\,ckpc from the QSO. 
The presence of one LAE with such small separation suggests a physical interaction with the QSO's host galaxy.
Intriguingly, the detected LAE and the possible \lya\ halo point to each other both in velocity and in position with respect to the QSO, opening the possibility that we may be witnessing an ongoing merger and that the \lya\ halo may be associated with the interaction among the LAE and the QSO's host galaxy.
We are cautious, however, that these results are based on the detection of a single LAE in proximity of one $z$$\sim$6.6 QSO.
Previous studies \citep[][]{Decarli2012,Banados2013,Mazzucchelli2016,Goto2017}, even if subjected to larger redshift uncertainties, did not report any evidence for an excess of LAEs in the proximity of $z$$\sim$6 QSOs \citep[but see][for the detection of \cii\ bright galaxies in proximity of $z$$\sim$6 QSOs]{Decarli2017}.
Measurements of the high redshift QSO--galaxy clustering in a statistical fashion is thus fundamental to discern among different scenarios proposed to explain the rapid formation of SMBHs at the end of the Cosmic reionization. 

\subsubsection{Physical Properties of the LAE}

The \lya\ emission of the companion galaxy could be boosted by the local enhancement of the ionizing background in the vicinity of a QSO.
Star formation can power the \lya\ emission up to a rest--frame equivalent width of W$_0$(\lya)$=$240\,\AA\ \citep[e.g.][]{Schaerer2002}. 
This value is commonly used as limit to identify LAEs associated with a fluorescent reprocessing of QSO radiation \citep[e.g.][]{Cantalupo2012}.
Some authors consider a less stringent limit of W$_0$(\lya)$>$100\,\AA\ for their selection \citep[e.g.][]{Trainor2013, Borisova2015}.
However, this may lead to a high level of contamination from non--fluorescent objects \citep[e.g.][]{Borisova2015}.

Here we test if the properties of the identified LAE are consistent with a \lya\ fluorescence scenario.
No significant continuum emission is detected redward of the \lya\ emission.
To constrain W$_0$(\lya) we thus consider the \mbox{1--$\sigma$} limit on the mean continuum ($\sigma_{\rm C}$) obtained averaging down the errors on the extracted spectrum ($\sigma_{\rm LAE,\lambda}$) over the wavelength range between 9281\,\AA\ and 9350\,\AA\ (i.e, up to the edge of the datacube): $\sigma^2_{\rm C}$=$\sum^{9350\mathring{\rm A}}_{\lambda=9281\mathring{\rm A}}{1/\sigma^2_{\rm LAE,\lambda}}$.
This leads to a \mbox{3--$\sigma$} lower limit on the rest--frame equivalent width of W$_{\rm 3\sigma}$(\lya)=$\frac{{\rm F}({\rm Ly}\alpha)}{3\,\sigma_{\rm C}(1+z_{\rm LAE})}$=7.4\,\AA, that is not strict enough to rule out a non--fluorescent scenario.
As a matter of fact, if powered by fluorescence, the luminosity of the LAE is proportional to ${\rm L}_{\nu_{\rm LL}}$ and inversely proportional to the square of the perpendicular distance from the QSO, assuming, for the sake of simplicity, the line--of--sight separation as negligible \citep[see][]{Hennawi2013}.
We would expect $\llya$$\sim$7$\times$10$^{43}$\,erg\,s$^{-1}$, where we considered the LAE as unresolved, i.e. with a diameter matching the seeing FWHM (0\farcs58).
The observed luminosity is a factor $\sim$30$\times$ fainter.
Note that this estimate depends on unknown quantities, namely the real distance of from the QSO, the luminosity of the QSO in the direction of the LAE, and the size of the surface illuminated by the ionizing radiation.
These could contribute to reduce the expected fluorescence emission.
On the other hand, it is possible that we are underestimating the real \lya\ luminosity.
Up to 90\% of the flux may come from an undetected extended component \citep[][]{Wisotzki2016} and a fraction of the \lya\ emission could be concealed by the neutral hydrogen.
Whitin these uncertainties, we favour a scenario where star formation, rather then fluorescence, is inducing the \lya\ emission.
From Equation~\ref{eq:sfr} we can thus derive the star--formation rate of the LAE: SFR$_{\rm LAE}$$\sim$1.3\,M$_\odot$\,yr$^{-1}$ \citep[see discussion in][ for the possibility of QSO feedback triggering the star formation in a closeby LAE at $z$$\sim$3.0]{Rauch2013}.

\section{SUMMARY AND CONCLUSIONS} \label{sec:summary}

In this paper we presented a sensitive search for extended \lya\ emission around the starbursting QSO \J0305\ at $z$$\sim$6.61.
The nominal \mbox{5--$\sigma$} surface brightness limit reached with MUSE is SB$_{5\sigma}^1$=1.9$\times$10$^{-18}$\,erg\,s$^{-1}$\,cm$^{-2}$\,arcsec$^{-2}$ over a 1\,arcsec$^2$ aperture (estimated collapsing the datacube over 5~wavelength slices centered at $\lambda$=9256\,\AA).
This formally corresponds to a luminosity limit of L$_{5\sigma}$$\sim$10$^{42.0}$\,erg\,s$^{-1}$ for unresolved sources with FWHM=200\,km\,s$^{-1}$, and to L$_{5\sigma}$$\sim$10$^{43.2}$\,erg\,s$^{-1}$ for a circular source with diameter 50\,kpc and the same FWHM.
The primary results of this study are:
\begin{itemize}
\item[1.] 
After carefully subtracting the unresolved emission from the central QSO we detect the presence of a tenuous [$\llya$=(3.0$\pm$0.4)$\times$10$^{42}$ erg\,s$^{-1}$] \lya\ halo extended over $\sim$10\,kpc.
To date, this is the first such nebula observed at $z$$>$6.5 and one of the faintest extended emission ever observed around a QSO at any redshift (see Figure~\ref{fig:red_lum}).
Despite the depth of our data, we do not detect the large scale (10--100\,kpc) \lya\ emission frequently observed in $z$$\sim$2--4 bright QSOs.
\item[2.]
A comparison between the \lya\ emission revealed by MUSE and the FIR properties of the host galaxy inferred from ALMA observations allows us to speculate on the geometry of the dust surrounding the QSO.
In particular, a patchy geometry of the dust cocoon can explain:
{\it (i)} The displacement between extended \lya\ and resolved [\cii] emission lines; 
{\it (ii)} the discrepancy between the observed \lya\ luminosity and expectation from the intense star formation of the host galaxy;
and {\it (iii)} the mild dust extinction present along the QSO line--of--sight.
This configuration permits the ionizing radiation from the newly formed stars to escape the host galaxy and give rise to the observed \lya\ emission.
\item[3.]
We estimate that the extended \lya\ emission is too faint to arise from recombinations on the ``skin'' of optically thick clouds.
A more plausible scenario is that the QSO radiation is sufficiently intense to maintain the surrounding gas highly ionized, hence we are observing fluorescent emission coming from optically thin clouds.
Intriguingly, a consequence of this emission mechanism is a hydrogen volume density of the gas illuminated by the QSO of $n_{\rm H}$$\sim$6\,cm$^{-3}$, similar to what estimated for the giant ($\gg$100\,kpc) \lya\ nebulae recently discovered around $z$$\sim$2 QSOsaffected.
\item[4.] 
A LAE with $\llya$$\sim$2.1$\times$10$^{42}$\,erg\,s$^{-1}$ is present at 12.5\,kpc and 560\,km\,s$^{-1}$ from \J0305.
Our current constraints on the rest--frame equivalent width [W$_0$(\lya)$>$7.4\,\AA] and luminosity, although not conclusive, disfavour a picture where the \lya\ emission is boosted by the QSO radiation.
Assuming that the \lya\ line is powered solely by star formation we derive a star--formation rate of SFR$_{\rm LAE}$$\sim$1.3\,M$_\odot$\,yr$^{-1}$.
\item[5.]
We calculate the enhanced probability of finding such a close LAE due to the clustering of galaxies around QSOs.
From the extrapolation of the $z$$\sim$6.6 QSO--LAE large--scale correlation function, we estimate this probability to be small ($<$10\%).
This supports a picture in which dissipative interaction and QSO activity in the young Universe are connected.
\end{itemize}

\acknowledgments

EPF, BPV, and FW acknowledge funding through the ERC grant `Cosmic Dawn'.
Support for RD was provided by the DFG priority program 1573 `The physics of the interstellar medium'.
EPF is grateful to M.~Fouesneau for providing support in the use of \textsc{Python} for the analysis of the MUSE datacubes, to M.~Fumagalli for sharing data on the QSO Q0956+122, and to M.~Rauch for providing feedback on the manuscript.
We thank the members of the ENIGMA group\footnote{\texttt{http://enigma.physics.ucsb.edu/}} at the Max Planck Institute for Astronomy (MPIA) for helpful discussions.
This research made use of \textsc{Astropy}, a community--developed core \textsc{Python} package for Astronomy \citep{Astropy2013}, of \textsc{APLpy}\footnote{\texttt{http://aplpy.github.io/}}, an open-source plotting package for \textsc{Python} based on \textsc{Matplotlib} \citep{Hunter2007}, and of \textsc{IRAF}\footnote{\textsc{IRAF} \citep{Tody1986, Tody1993}, is distributed by the National Optical Astronomy Observatories, which are operated by the Association of Universities for Research in Astronomy, Inc., under cooperative agreement with the National Science Foundation.}.
Based on observations collected at the European Organisation for Astronomical Research in the Southern Hemisphere under ESO programme 094.B-0893(A).


\begin{thebibliography}{}

\expandafter\ifx\csname natexlab\endcsname\relax\def\natexlab#1{#1}\fi

\bibitem[Adams et al.(2015)]{Adams2015} Adams, S.~M., Martini, P., Croxall, K.~V., Overzier, R.~A., \& Silverman, J.~D.\ 2015, \mnras, 448, 1335 
\bibitem[Alam \& Miralda-Escud{\'e}(2002)]{Alam2002} Alam, S.~M.~K., \& Miralda-Escud{\'e}, J.\ 2002, \apj, 568, 576 
\bibitem[Aravena et al.(2016)]{Aravena2016} Aravena, M., Decarli, R., Walter, F., et al.\ 2016, arXiv:1607.06769 
\bibitem[Arrigoni Battaia et al.(2015{\natexlab{a}})]{Arrigoni2015} Arrigoni Battaia, F., Yang, Y., Hennawi, J.~F., et al.\ 2015, \apj, 804, 26 
\bibitem[Arrigoni Battaia et al.(2015{\natexlab{b}})]{Arrigoni2015giant} Arrigoni Battaia, F., Hennawi, J.~F., Prochaska, J.~X., \& Cantalupo, S.\ 2015, \apj, 809, 163 
\bibitem[Arrigoni Battaia et al.(2016)]{Arrigoni2016} Arrigoni Battaia, F., Hennawi, J.~F., Cantalupo, S., \& Prochaska, J.~X.\ 2016, \apj, 829, 3
\bibitem[Astropy Collaboration et al.(2013)]{Astropy2013} Astropy Collaboration, Robitaille, T.~P., Tollerud, E.~J., et al.\ 2013, \aap, 558, A33 
\bibitem[Bacon et al.(2010)]{Bacon2010} Bacon, R., Accardo, M., Adjali, L., et al.\ 2010, \procspie, 7735, 773508
\bibitem[Bacon et al.(2015)]{Bacon2015} Bacon, R., Brinchmann, J., Richard, J., et al.\ 2015, \aap, 575, A75 
\bibitem[Ba{\~n}ados et al.(2013)]{Banados2013} Ba{\~n}ados, E., Venemans, B., Walter, F., et al.\ 2013, \apj, 773, 178 
\bibitem[Ba{\~n}ados et al.(2014)]{Banados2014} Ba{\~n}ados, E., Venemans, B.~P., Morganson, E., et al.\ 2014, \aj, 148, 14
\bibitem[Ba{\~n}ados et al.(2015{\natexlab{a}})]{Banados2015RL} Ba{\~n}ados, E., Venemans, B.~P., Morganson, E., et al.\ 2015, \apj, 804, 118 
\bibitem[Ba{\~n}ados et al.(2015{\natexlab{b}})]{Banados2015} Ba{\~n}ados, E., Decarli, R., Walter, F., et al.\ 2015, \apjl, 805, L8 
\bibitem[Ba{\~n}ados et al.(2016)]{Banados2016} Ba{\~n}ados, E., Venemans, B.~P., Decarli, R., et al.\ 2016, arXiv:1608.03279 
\bibitem[Barnett et al.(2015)]{Barnett2015} Barnett, R., Warren, S.~J., Banerji, M., et al.\ 2015, \aap, 575, A31 
\bibitem[Beelen et al.(2006)]{Beelen2006} Beelen, A., Cox, P., Benford, D.~J., et al.\ 2006, \apj, 642, 694 
\bibitem[Bertin \& Arnouts(1996)]{Bertin1996} Bertin, E., \& Arnouts, S.\ 1996, \aaps, 117, 393 
\bibitem[Bergeron et al.(1999)]{Bergeron1999} Bergeron, J., Petitjean, P., Cristiani, S., et al.\ 1999, \aap, 343, L40 
\bibitem[Bonning et al.(2007)]{Bonning2007} Bonning, E.~W., Shields, G.~A., \& Salviander, S.\ 2007, \apjl, 666, L13 
\bibitem[Borisova et al.(2015)]{Borisova2015} Borisova, E., Lilly, S.~J., Cantalupo, S., et al.\ 2015, arXiv:1510.00029 
\bibitem[Borisova et al.(2016)]{Borisova2016} Borisova, E., Cantalupo, S., Lilly, S.~J., et al.\ 2016, arXiv:1605.01422 
\bibitem[Bowen et al.(2006)]{Bowen2006} Bowen, D.~V., Hennawi, J.~F., M{\'e}nard, B., et al.\ 2006, \apjl, 645, L105 
\bibitem[Barrio et al.(2008)]{Barrio2008} Barrio, F.~E., Jarvis, M.~J., Rawlings, S., et al.\ 2008, \mnras, 389, 792 
\bibitem[Bremer et al.(1992)]{Bremer1992} Bremer, M.~N., Fabian, A.~C., Sargent, W.~L.~W., et al.\ 1992, \mnras, 258, 23P 
\bibitem[Bunker et al.(2003)]{Bunker2003} Bunker, A., Smith, J., Spinrad, H., Stern, D., \& Warren, S.\ 2003, \apss, 284, 357 
\bibitem[Cantalupo et al.(2005)]{Cantalupo2005} Cantalupo, S., Porciani, C., Lilly, S.~J., \& Miniati, F.\ 2005, \apj, 628, 61 
\bibitem[Casey et al.(2014)]{Casey2014} Casey, C.~M., Scoville, N.~Z., Sanders, D.~B., et al.\ 2014, \apj, 796, 95 
\bibitem[Cantalupo et al.(2012)]{Cantalupo2012} Cantalupo, S., Lilly, S.~J., \& Haehnelt, M.~G.\ 2012, \mnras, 425, 1992 
\bibitem[Cantalupo et al.(2014)]{Cantalupo2014} Cantalupo, S., Arrigoni-Battaia, F., Prochaska, J.~X., Hennawi, J.~F., \& Madau, P.\ 2014, \nat, 506, 63 
\bibitem[Carnall et al.(2015)]{Carnall2015} Carnall, A.~C., Shanks, T., Chehade, B., et al.\ 2015, \mnras, 451, L16 
\bibitem[Christensen et al.(2006)]{Christensen2006} Christensen, L., Jahnke, K., Wisotzki, L., \& S{\'a}nchez, S.~F.\ 2006, \aap, 459, 717 
\bibitem[Courbin et al.(2008)]{Courbin2008} Courbin, F., North, P., Eigenbrod, A., \& Chelouche, D.\ 2008, \aap, 488, 91 
\bibitem[De Breuck et al.(2004)]{DeBreuck2004} De Breuck, C., Bertoldi, F., Carilli, C., et al.\ 2004, \aap, 424, 1 
\bibitem[da Cunha et al.(2013)]{daCunha2013} da Cunha, E., Groves, B., Walter, F., et al.\ 2013, \apj, 766, 13 
\bibitem[Decarli et al.(2009)]{Decarli2009} Decarli, R., Treves, A., \& Falomo, R.\ 2009, \mnras, 396, L31 
\bibitem[Decarli et al.(2012)]{Decarli2012} Decarli, R., Walter, F., Yang, Y., et al.\ 2012, \apj, 756, 150
\bibitem[Decarli et al.(2016)]{Decarli2016} Decarli, R., Walter, F., Aravena, M., et al.\ 2016, arXiv:1607.06771
\bibitem[Decarli et al.(2017)]{Decarli2017} Decarli, R., Walter, F., Venemans, B.~P., et al.\ 2017, \nat, 545, 457 
\bibitem[De Rosa et al.(2014)]{Derosa2014} De Rosa, G., Venemans, B.~P., Decarli, R., et al.\ 2014, \apj, 790, 145 
\bibitem[Di Matteo et al.(2012)]{Dimatteo2012} Di Matteo, T., Khandai, N., DeGraf, C., et al.\ 2012, \apjl, 745, L29 
\bibitem[Djorgovski(1991)]{Djorgovski1991} Djorgovski, S.\ 1991, The Space Distribution of Quasars, 21, 349 
\bibitem[Djorgovski et al.(2007)]{Djorgovski2007} Djorgovski, S.~G., Courbin, F., Meylan, G., et al.\ 2007, \apjl, 662, L1 
\bibitem[Dubois et al.(2012)]{Dubois2012} Dubois, Y., Pichon, C., Haehnelt, M., et al.\ 2012, \mnras, 423, 3616 
\bibitem[Edge et al.(2013)]{Edge2013} Edge, A., Sutherland, W., Kuijken, K., et al.\ 2013, The Messenger, 154, 32 
\bibitem[Eftekharzadeh et al.(2015)]{Eftekharzadeh2015} Eftekharzadeh, S., Myers, A.~D., White, M., et al.\ 2015, \mnras, 453, 2779 
\bibitem[Eftekharzadeh et al.(2017)]{Eftekharzadeh2017} Eftekharzadeh, S., Myers, A.~D., Hennawi, J.~F., et al.\ 2017, \mnras, 468, 77 
\bibitem[Fan et al.(2006)]{Fan2006} Fan, X., Strauss, M.~A., Richards, G.~T., et al.\ 2006, \aj, 131, 1203 
\bibitem[Farina et al.(2011)]{Farina2011} Farina, E.~P., Falomo, R., \& Treves, A.\ 2011, \mnras, 415, 3163 
\bibitem[Farina et al.(2013{\natexlab{a}})]{Farina2013} Farina, E.~P., Falomo, R., Decarli, R., Treves, A., \& Kotilainen, J.~K.\ 2013, \mnras, 429, 1267 
\bibitem[Farina et al.(2013{\natexlab{b}})]{Farina2013trip} Farina, E.~P., Montuori, C., Decarli, R., \& Fumagalli, M.\ 2013, \mnras, 431, 1019 
\bibitem[Farina et al.(2014)]{Farina2014} Farina, E.~P., Falomo, R., Scarpa, R., et al.\ 2014, \mnras, 441, 886 
\bibitem[Fathivavsari et al.(2016)]{Fathivavsari2016} Fathivavsari, H., Petitjean, P., Noterdaeme, P., et al.\ 2016, \mnras, 461, 1816 
\bibitem[Feng et al.(2014)]{Feng2014} Feng, Y., Di Matteo, T., Croft, R., \& Khandai, N.\ 2014, \mnras, 440, 1865 
\bibitem[Francis \& McDonnell(2006)]{Francis2006} Francis, P.~J., \& McDonnell, S.\ 2006, \mnras, 370, 1372 
\bibitem[Fumagalli et al.(2016)]{Fumagalli2016} Fumagalli, M., Cantalupo, S., Dekel, A., et al.\ 2016, arXiv:1607.03893 
\bibitem[Fynbo et al.(2000)]{Fynbo2000} Fynbo, J.~U., Thomsen, B., \& M{\o}ller, P.\ 2000, \aap, 353, 457 
\bibitem[Garcia-Vergara et al.(2017)]{Garcia2017} Garcia-Vergara, C., Hennawi, J.~F., Barrientos, L.~F., \& Rix, H.-W.\ 2017, arXiv:1701.01114 
\bibitem[Gallerani et al.(2010)]{Gallerani2010} Gallerani, S., Maiolino, R., Juarez, Y., et al.\ 2010, \aap, 523, A85 
\bibitem[Gehrels(1986)]{Gehrels1986} Gehrels, N.\ 1986, \apj, 303, 336 
\bibitem[Goto et al.(2009)]{Goto2009} Goto, T., Utsumi, Y., Furusawa, H., Miyazaki, S., \& Komiyama, Y.\ 2009, \mnras, 400, 843 
\bibitem[Goto et al.(2012)]{Goto2012} Goto, T., Utsumi, Y., Walsh, J.~R., et al.\ 2012, \mnras, 421, L77
\bibitem[Goto et al.(2017)]{Goto2017} Goto, T., Utsumi, Y., Kikuta, S., et al.\ 2017, arXiv:1706.04620 
\bibitem[Gordon et al.(2003)]{Gordon2003} Gordon, K.~D., Clayton, G.~C., Misselt, K.~A., Landolt, A.~U., \& Wolff, M.~J.\ 2003, \apj, 594, 279 
\bibitem[Gould \& Weinberg(1996)]{Gould1996} Gould, A., \& Weinberg, D.~H.\ 1996, \apj, 468, 462 
\bibitem[Haiman \& Rees(2001)]{Haiman2001} Haiman, Z., \& Rees, M.~J.\ 2001, \apj, 556, 87 
\bibitem[Hashimoto et al.(2013)]{Hashimoto2013} Hashimoto, T., Ouchi, M., Shimasaku, K., et al.\ 2013, \apj, 765, 70 
\bibitem[Heckman et al.(1991{\natexlab{a}})]{Heckman1991a} Heckman, T.~M., Miley, G.~K., Lehnert, M.~D., \& van Breugel, W.\ 1991, \apj, 370, 78
\bibitem[Heckman et al.(1991{\natexlab{b}})]{Heckman1991b} Heckman, T.~M., Lehnert, M.~D., Miley, G.~K., \& van Breugel, W.\ 1991, \apj, 381, 373 
\bibitem[Hennawi et al.(2006{\natexlab{a}})]{Hennawi2006} Hennawi, J.~F., Prochaska, J.~X., Burles, S., et al.\ 2006, \apj, 651, 61 
\bibitem[Hennawi et al.(2006{\natexlab{b}})]{Hennawi2006b} Hennawi, J.~F., Strauss, M.~A., Oguri, M., et al.\ 2006, \aj, 131, 1 
\bibitem[Hennawi \& Prochaska(2007)]{Hennawi2007} Hennawi, J.~F., \& Prochaska, J.~X.\ 2007, \apj, 655, 735 
\bibitem[Hennawi et al.(2010)]{Hennawi2010} Hennawi, J.~F., Myers, A.~D., Shen, Y., et al.\ 2010, \apj, 719, 1672 
\bibitem[Hennawi \& Prochaska(2013)]{Hennawi2013} Hennawi, J.~F., \& Prochaska, J.~X.\ 2013, \apj, 766, 58 
\bibitem[Hennawi et al.(2015)]{Hennawi2015} Hennawi, J.~F., Prochaska, J.~X., Cantalupo, S., \& Arrigoni-Battaia, F.\ 2015, Science, 348, 779 
\bibitem[Herenz et al.(2015)]{Herenz2015} Herenz, E.~C., Wisotzki, L., Roth, M., \& Anders, F.\ 2015, \aap, 576, A115 
\bibitem[Hewett \& Wild(2010)]{Hewett2010} Hewett, P.~C., \& Wild, V.\ 2010, \mnras, 405, 2302 
\bibitem[Hodge et al.(2015)]{Hodge2015} Hodge, J.~A., Riechers, D., Decarli, R., et al.\ 2015, \apjl, 798, L18 
\bibitem[Hu et al.(2010)]{Hu2010} Hu, E.~M., Cowie, L.~L., Barger, A.~J., et al.\ 2010, \apj, 725, 394 
\bibitem[Humphrey et al.(2013)]{Humphrey2013} Humphrey, A., Binette, L., Villar-Mart{\'{\i}}n, M., Aretxaga, I., \& Papaderos, P.\ 2013, \mnras, 428, 563
\bibitem[Hunter (2007)]{Hunter2007} Hunter, J. D.\ 2007, Computing in Science and Engineering, 9, 90 8
\bibitem[Husband et al.(2015)]{Husband2015} Husband, K., Bremer, M.~N., Stanway, E.~R., \& Lehnert, M.~D.\ 2015, \mnras, 452, 2388 
\bibitem[Jiang et al.(2009)]{Jiang2009} Jiang, L., Fan, X., Bian, F., et al.\ 2009, \aj, 138, 305 
\bibitem[Jiang et al.(2016)]{Jiang2016} Jiang, L., McGreer, I.~D., Fan, X., et al.\ 2016, arXiv:1610.05369 
\bibitem[Johnson et al.(2015)]{Johnson2015} Johnson, S.~D., Chen, H.-W., \& Mulchaey, J.~S.\ 2015, \mnras, 452, 2553 
\bibitem[Kayo \& Oguri(2012)]{Kayo2012} Kayo, I., \& Oguri, M.\ 2012, \mnras, 424, 1363 
\bibitem[Karhunen et al.(2014)]{Karhunen2014} Karhunen, K., Kotilainen, J.~K., Falomo, R., \& Bettoni, D.\ 2014, \mnras, 441, 1802 
\bibitem[Kennicutt \& Evans(2012)]{Kennicutt2012} Kennicutt, R.~C., \& Evans, N.~J.\ 2012, \araa, 50, 531 
\bibitem[Lau et al.(2016)]{Lau2016} Lau, M.~W., Prochaska, J.~X., \& Hennawi, J.~F.\ 2016, \apjs, 226, 25
\bibitem[Lau et al.(2017)]{Lau2017} Lau, M.~W., Prochaska, J.~X., \& Hennawi, J.~F.\ 2017, arXiv:1705.03476 
\bibitem[Lehnert \& Becker(1998)]{Lehnert1998} Lehnert, M.~D., \& Becker, R.~H.\ 1998, \aap, 332, 514 
\bibitem[Leipski et al.(2014)]{Leipski2014} Leipski, C., Meisenheimer, K., Walter, F., et al.\ 2014, \apj, 785, 154 
\bibitem[Lusso et al.(2013)]{Lusso2013} Lusso, E., Hennawi, J.~F., Comastri, A., et al.\ 2013, \apj, 777, 86 
\bibitem[Lusso et al.(2015)]{Lusso2015} Lusso, E., Worseck, G., Hennawi, J.~F., et al.\ 2015, \mnras, 449, 4204 
\bibitem[Madau et al.(2014)]{Madau2014} Madau, P., Haardt, F., \& Dotti, M.\ 2014, \apjl, 784, L38 
\bibitem[Martin et al.(2014)]{Martin2014} Martin, D.~C., Chang, D., Matuszewski, M., et al.\ 2014, \apj, 786, 106 
\bibitem[Matsuda et al.(2011)]{Matsuda2011} Matsuda, Y., Yamada, T., Hayashino, T., et al.\ 2011, \mnras, 410, L13 
\bibitem[Matsuoka et al.(2016)]{Matsuoka2016} Matsuoka, Y., Onoue, M., Kashikawa, N., et al.\ 2016, arXiv:1603.02281 
\bibitem[Matsuoka et al.(2017)]{Matsuoka2017} Matsuoka, Y., Onoue, M., Kashikawa, N., et al.\ 2017, arXiv:1704.05854 
\bibitem[Matthee et al.(2015)]{Matthee2015} Matthee, J., Sobral, D., Santos, S., et al.\ 2015, \mnras, 451, 400 
\bibitem[Mazzucchelli et al.(2016)]{Mazzucchelli2016} Mazzucchelli, C., Ba{\~n}ados, E., Decarli, R., et al.\ 2016, arXiv:1611.02870 
\bibitem[Mazzucchelli et al.(2017)]{Mazzucchelli2017} Mazzucchelli, C., et al.\ 2017, Submitted 
\bibitem[Merloni et al.(2014)]{Merloni2014} Merloni, A., Bongiorno, A., Brusa, M., et al.\ 2014, \mnras, 437, 3550 
\bibitem[McGreer et al.(2014)]{McGreer2014} McGreer, I.~D., Fan, X., Strauss, M.~A., et al.\ 2014, \aj, 148, 73 
\bibitem[McGreer et al.(2016)]{McGreer2016} McGreer, I.~D., Eftekharzadeh, S., Myers, A.~D., \& Fan, X.\ 2016, \aj, 151, 61 
\bibitem[Meiksin(2006)]{Meiksin2006} Meiksin, A.\ 2006, \mnras, 365, 807 
\bibitem[Mechtley et al.(2012)]{Mechtley2012} Mechtley, M., Windhorst, R.~A., Ryan, R.~E., et al.\ 2012, \apjl, 756, L38 
\bibitem[Myers et al.(2007{\natexlab{a}})]{Myers2007a} Myers, A.~D., Brunner, R.~J., Nichol, R.~C., et al.\ 2007, \apj, 658, 85 
\bibitem[Myers et al.(2007{\natexlab{b}})]{Myers2007b} Myers, A.~D., Brunner, R.~J., Richards, G.~T., et al.\ 2007, \apj, 658, 99 
\bibitem[Myers et al.(2008)]{Myers2008} Myers, A.~D., Richards, G.~T., Brunner, R.~J., et al.\ 2008, \apj, 678, 635-646 
\bibitem[Morrissey et al.(2012)]{Morrissey2012} Morrissey, P., Matuszewski, M., Martin, C., et al.\ 2012, \procspie, 8446, 844613 
\bibitem[Mortlock et al.(2011)]{Mortlock2011} Mortlock, D.~J., Warren, S.~J., Venemans, B.~P., et al.\ 2011, \nat, 474, 616 
\bibitem[North et al.(2012)]{North2012} North, P.~L., Courbin, F., Eigenbrod, A., \& Chelouche, D.\ 2012, \aap, 542, A91 
\bibitem[Ouchi et al.(2003)]{Ouchi2003} Ouchi, M., Shimasaku, K., Furusawa, H., et al.\ 2003, \apj, 582, 60 
\bibitem[Ouchi et al.(2010)]{Ouchi2010} Ouchi, M., Shimasaku, K., Furusawa, H., et al.\ 2010, \apj, 723, 869 
\bibitem[Padmanabhan et al.(2009)]{Padmanabhan2009} Padmanabhan, N., White, M., Norberg, P., \& Porciani, C.\ 2009, \mnras, 397, 1862 
\bibitem[Peebles(1980)]{Peebles1980} Peebles, P.~J.~E.\ 1980, Research supported by the National Science Foundation.~Princeton, N.J., Princeton University Press, 1980.~435 p.,
\bibitem[Pentericci et al.(2016)]{Pentericci2016} Pentericci, L., Carniani, S., Castellano, M., et al.\ 2016, \apjl, 829, L11 
\bibitem[Prochaska \& Hennawi(2009)]{Prochaska2009} Prochaska, J.~X., \& Hennawi, J.~F.\ 2009, \apj, 690, 1558 
\bibitem[Prochaska et al.(2013{\natexlab{a}})]{Prochaska2013a} Prochaska, J.~X., Hennawi, J.~F., Lee, K.-G., et al.\ 2013, \apj, 776, 136 
\bibitem[Prochaska et al.(2013{\natexlab{b}})]{Prochaska2013b} Prochaska, J.~X., Hennawi, J.~F., \& Simcoe, R.~A.\ 2013, \apjl, 762, L19 
\bibitem[Prochaska et al.(2014)]{Prochaska2014} Prochaska, J.~X., Lau, M.~W., \& Hennawi, J.~F.\ 2014, \apj, 796, 140 
\bibitem[Rauch et al.(2013)]{Rauch2013} Rauch, M., Becker, G.~D., Haehnelt, M.~G., Carswell, R.~F., \& Gauthier, J.-R.\ 2013, \mnras, 431, L68 
\bibitem[Reed et al.(2015)]{Reed2015} Reed, S.~L., McMahon, R.~G., Banerji, M., et al.\ 2015, \mnras, 454, 3952 
\bibitem[Rees(1988)]{Rees1988} Rees, M.~J.\ 1988, \mnras, 231, 91p 
\bibitem[Richards et al.(2002)]{Richards2002} Richards, G.~T., Vanden Berk, D.~E., Reichard, T.~A., et al.\ 2002, \aj, 124, 1 
\bibitem[Richardson et al.(2012)]{Richardson2012} Richardson, J., Zheng, Z., Chatterjee, S., Nagai, D., \& Shen, Y.\ 2012, \apj, 755, 30 
\bibitem[Roche et al.(2014)]{Roche2014} Roche, N., Humphrey, A., \& Binette, L.\ 2014, \mnras, 443, 3795 
\bibitem[Roettgering et al.(1997)]{Roettgering1997} Roettgering, H.~J.~A., van Ojik, R., Miley, G.~K., et al.\ 1997, \aap, 326, 505 
\bibitem[Ross et al.(2009)]{Ross2009} Ross, N.~P., Shen, Y., Strauss, M.~A., et al.\ 2009, \apj, 697, 1634 
\bibitem[Sandrinelli et al.(2014)]{Sandrinelli2014} Sandrinelli, A., Falomo, R., Treves, A., Farina, E.~P., \& Uslenghi, M.\ 2014, \mnras, 444, 1835 
\bibitem[Schaerer(2002)]{Schaerer2002} Schaerer, D.\ 2002, \aap, 382, 28 
\bibitem[Schneider(2006)]{Schneider2006} Schneider, P.\ 2006, Extragalactic Astronomy and Cosmology, by Peter Schneider.~Berlin: Springer, 2006.,  
\bibitem[Selsing et al.(2016)]{Selsing2016} Selsing, J., Fynbo, J.~P.~U., Christensen, L., \& Krogager, J.-K.\ 2016, \aap, 585, A87 
\bibitem[Serber et al.(2006)]{Serber2006} Serber, W., Bahcall, N., M{\'e}nard, B., \& Richards, G.\ 2006, \apj, 643, 68 
\bibitem[Shen et al.(2007)]{Shen2007} Shen, Y., Strauss, M.~A., Oguri, M., et al.\ 2007, \aj, 133, 2222 
\bibitem[Shen et al.(2010)]{Shen2010} Shen, Y., Hennawi, J.~F., Shankar, F., et al.\ 2010, \apj, 719, 1693 
\bibitem[Shen et al.(2013)]{Shen2013} Shen, Y., McBride, C.~K., White, M., et al.\ 2013, \apj, 778, 98 
\bibitem[Smith et al.(2009)]{Smith2009} Smith, D.~J.~B., Jarvis, M.~J., Simpson, C., \& Mart{\'{\i}}nez-Sansigre, A.\ 2009, \mnras, 393, 309 
\bibitem[Stark et al.(2015)]{Stark2015} Stark, D.~P., Richard, J., Charlot, S., et al.\ 2015, \mnras, 450, 1846 
\bibitem[Steidel et al.(2010)]{Steidel2010} Steidel, C.~C., Erb, D.~K., Shapley, A.~E., et al.\ 2010, \apj, 717, 289 
\bibitem[Swinbank et al.(2015)]{Swinbank2015} Swinbank, A.~M., Vernet, J.~D.~R., Smail, I., et al.\ 2015, \mnras, 449, 1298 
\bibitem[Tody(1986)]{Tody1986} Tody, D.\ 1986, \procspie, 627, 733 
\bibitem[Tody(1993)]{Tody1993} Tody, D.\ 1993, Astronomical Data Analysis Software and Systems II, 52, 173 
\bibitem[Trainor \& Steidel(2013)]{Trainor2013} Trainor, R., \& Steidel, C.~C.\ 2013, \apjl, 775, L3 
\bibitem[Treister et al.(2008)]{Treister2008} Treister, E., Krolik, J.~H., \& Dullemond, C.\ 2008, \apj, 679, 140-148 
\bibitem[Valiante et al.(2011)]{Valiante2011} Valiante, R., Schneider, R., Salvadori, S., \& Bianchi, S.\ 2011, \mnras, 416, 1916 
\bibitem[van Ojik et al.(1997)]{vanOjik1997} van Ojik, R., Roettgering, H.~J.~A., Miley, G.~K., \& Hunstead, R.~W.\ 1997, \aap, 317, 358 
\bibitem[Venemans et al.(2007)]{Venemans2007} Venemans, B.~P., R{\"o}ttgering, H.~J.~A., Miley, G.~K., et al.\ 2007, \aap, 461, 823 
\bibitem[Venemans et al.(2012)]{Venemans2012} Venemans, B.~P., McMahon, R.~G., Walter, F., et al.\ 2012, \apjl, 751, L25 
\bibitem[Venemans et al.(2013)]{Venemans2013} Venemans, B.~P., Findlay, J.~R., Sutherland, W.~J., et al.\ 2013, \apj, 779, 24 
\bibitem[Venemans et al.(2015{\natexlab{a}})]{Venemans2015lowz} Venemans, B.~P., Verdoes Kleijn, G.~A., Mwebaze, J., et al.\ 2015, \mnras, 453, 2259
\bibitem[Venemans et al.(2015{\natexlab{b}})]{Venemans2015highz} Venemans, B.~P., Ba{\~n}ados, E., Decarli, R., et al.\ 2015, \apjl, 801, L11 
\bibitem[Venemans et al.(2016)]{Venemans2016} Venemans, B.~P., Walter, F., Zschaechner, L., et al.\ 2016, \apj, 816, 37 
\bibitem[Volonteri \& Rees(2005)]{Volonteri2005} Volonteri, M., \& Rees, M.~J.\ 2005, \apj, 633, 624 
\bibitem[Volonteri(2010)]{Volonteri2010} Volonteri, M.\ 2010, \aapr, 18, 279 
\bibitem[Volonteri(2012)]{Volonteri2012} Volonteri, M.\ 2012, Science, 337, 544 
\bibitem[Volonteri et al.(2015)]{Volonteri2015} Volonteri, M., Silk, J., \& Dubus, G.\ 2015, \apj, 804, 148 
\bibitem[Weidinger et al.(2004)]{Weidinger2004} Weidinger, M., M{\o}ller, P., \& Fynbo, J.~P.~U.\ 2004, \nat, 430, 999 
\bibitem[Weidinger et al.(2005)]{Weidinger2005} Weidinger, M., M{\o}ller, P., Fynbo, J.~P.~U., \& Thomsen, B.\ 2005, \aap, 436, 825 
\bibitem[Weilbacher et al.(2012)]{Weilbacher2012} Weilbacher, P.~M., Streicher, O., Urrutia, T., et al.\ 2012, \procspie, 8451, 84510B 
\bibitem[Weilbacher et al.(2014)]{Weilbacher2014} Weilbacher, P.~M., Streicher, O., Urrutia, T., et al.\ 2014, Astronomical Data Analysis Software and Systems XXIII, 485, 451 
\bibitem[White et al.(2012)]{White2012} White, M., Myers, A.~D., Ross, N.~P., et al.\ 2012, \mnras, 424, 933 
\bibitem[Willott et al.(2010)]{Willott2010} Willott, C.~J., Delorme, P., Reyl{\'e}, C., et al.\ 2010, \aj, 139, 906 
\bibitem[Willott et al.(2011)]{Willott2011} Willott, C.~J., Chet, S., Bergeron, J., \& Hutchings, J.~B.\ 2011, \aj, 142, 186 
\bibitem[Wisotzki et al.(2016)]{Wisotzki2016} Wisotzki, L., Bacon, R., Blaizot, J., et al.\ 2016, \aap, 587, A98 
\bibitem[Yoo \& Miralda-Escud{\'e}(2004)]{Yoo2004} Yoo, J., \& Miralda-Escud{\'e}, J.\ 2004, \apjl, 614, L25 
\bibitem[Yang et al.(2009)]{Yang2009} Yang, Y., Zabludoff, A., Tremonti, C., Eisenstein, D., \& Dav{\'e}, R.\ 2009, \apj, 693, 1579 
\bibitem[Zhang et al.(2013)]{Zhang2013} Zhang, S., Wang, T., Wang, H., \& Zhou, H.\ 2013, \apj, 773, 175 
\bibitem[Zirm et al.(2005)]{Zirm2005} Zirm, A.~W., Overzier, R.~A., Miley, G.~K., et al.\ 2005, \apj, 630, 68 

\end{thebibliography}
\end{document}